\documentclass[twocolumn]{aastex631}
\usepackage{color}
\usepackage{makecell}
\usepackage{appendix}

\usepackage{amsmath}
\usepackage{mathrsfs}

\shorttitle{the possible neutrino emission in GRB 230307A }
\shortauthors{Xin-Ying Song}

\graphicspath{{./}{figures/}}

\begin{document}

\title{On the `Loose' Constraint from IceCube Neutrino Non-Detection of GRB 230307A}

\correspondingauthor{Xin-Ying Song}
\email{songxy@ihep.ac.cn}



\author[0000-0002-2176-8778]{Xin-Ying Song}
\affiliation{Key Laboratory of Particle Astrophysics, Institute of High Energy Physics, Chinese Academy of Sciences, Beijing 100049, China}

\begin{abstract}
The recent extremely bright gamma-ray burst (GRB), GRB 230307A from a binary neutron star merger may offer a good probe for the production of GRB-neutrinos. Within the constraint from IceCube neutrino non-detection, the limits for key physical parameters of this burst are extracted in different scenarios including the fireball, Poynting-flux-dominated (PFD) and hybrid jet. Different from the former nearby `monsters' and due to its smaller isotropic equivalent radiated energy ($E_{\gamma,\rm iso}\sim4\times10^{52}$ erg), the constraint seems loose if non-thermal neutrinos produced from photomeson interactions are the only consideration. However, a quasi-thermal neutrino emission from hadronuclear processes is constrained in this neutron-rich post-merger environment, and the upper limit of the allowed nucleon loading factor is about a few. Based on this, a discussion is presented on the possible prompt emission mechanism and jet composition for GRB 230307A in the context of multi-messenger astrophysics. \textbf{It is worth noting that till now no GRB-neutrinos have been ever detected, even for the two brightest nearby GRBs ever observed (GRB 221009A and GRB 230307A) which have different dissipation mechanisms.}

\end{abstract}



\section{Introduction} \label{sec:intro}
For decades, GRBs are proposed to be ultra-high-energy cosmic-ray (UHECR) accelerators and power potential sources of astrophysical high energy neutrinos in the GRB fireball scenario~\cite[e.g.][]{1994ApJ...427..708P,1995PhRvL..75..386W,1995ApJ...453..883V,2012PhRvL.108w1101H}. There are mainly two mechanisms for the production of the neutrinos relevant to GRBs: 1) the photomeson ($p\gamma$) process~\cite[e.g.][]{1997PhRvL..78.2292W}, in which the photons of GRB are guaranteed photons; as predicted in a model-dependent theory, the prompt emission sites are different, which could affect the fluence of this kind of non-thermal (NT) neutrinos~\cite[e.g.][]{2013PhRvL.110l1101Z}. 
2) Hadronuclear ($pn$, $pp$ and $nn$) interactions ~\cite[e.g.][]{2000PhRvL..85.1362B,2007A&A...471..395K, 2013PhRvL.111m1102M,2013PhRvL.110x1101B} which produce quasi-thermal (QT) neutrinos during the expanding of hot jet material of the outflow; neutrons could be decoupled from protons if the bulk Lorentz Factor ($\Gamma$) is large enough. Thus inelastic collisions could occur and pions could be produced. The neutrinos with 5-10 GeV could be generated after pions decays; alternatively, if the coasting occurs earlier than $np$ decoupling, the dissipation of neutrons via internal collisions
between the compound flows may happen, which could also generate QT neutrinos in the sub-TeV range~\citep[e.g.][]{2013PhRvL.111m1102M}. Note that the neutrino emission could occur during the prompt phase of GRB as well as the (choked) jet’s propagation
within the macronova/kilonova ejecta or a massive progenitor envelope ~\citep[e.g.][]{2001PhRvL..87q1102M,2013PhRvL.111l1102M,2016PhRvD..93h3003S,2018PhRvD..98d3020K}, as a precursor for GRBs. However, till now, no evident correlation between neutrino events and
observed GRBs is observed~\citep[e.g.][]{2017ApJ...843..112A,2023ApJ...951...45A}.

 It is usually proposed that long GRBs are from the collapses of massive stars, while short GRBs are produced from mergers of compact stars. However, there are some exceptions, such as GRB 060614~\citep[e.g.][]{2006Natur.444.1044G} and GRB 211211A~\cite[e.g.][]{2022Natur.612..232Y}. Recently, another exception, GRB 230307A, associated with kilonova emission~\citep[][]{2023GCN.33569....1L,2023GCN.33405....1F,2023GCN.33406....1X,2023GCN.33427....1S} arises.
 The isotropic-equivalent radiated energy $E_{\gamma, \rm iso}\sim 4\times10^{52}$ erg of GRB 230307A is smaller compared with other nearby bright GRBs (e.g., $E_{\gamma, \rm iso}\sim 10^{54}$ erg for GRB 130427A and $E_{\gamma, \rm iso}\sim2\times 10^{54}$ erg for GRB 221009A), thus it could be inferred that the features of the burst can not be well constrained if the neutrino emission from $p\gamma$ interactions is the only consideration.
 However, with the favoured redshift $z\sim0.065$~\citep[e.g.][]{2023GCN.33447....1O,2023arXiv230702098L}, $d^2_{L}$ of GRB 230307A is about 6 times smaller than that of GRB 221009A, and 37 times smaller than that of GRB 130427A, where $d_L$ is the distance of the GRB source. Therefore it may be still a good probe for the GRB-neutrinos. 

  It is worth to note that the IceCube Neutrino Observatory provides an upper limit (U.L.) of muon neutrinos ($\nu_{\mu}$), $1.0$ GeV cm$^{-2}$ at 90\% C.L.~\citep[][]{2023GCN.33430....1I}, which seems less constraining compared with $3.9\times10^{-2}$ GeV cm$^{-2}$ at 90\% CL of GRB 221009A~\cite[][]{2022GCN.32665....1I}.  Besides, there exists a significant photospheric emission in the prompt phase in GRB 230307A, which is different from synchrotron radiation-dominated GRB 221009A whose jet is dominated by the Poynting flux~\citep{2023ApJ...947L..11Y}. The emission mechanism as well as the jet composition could affect the production of the neutrinos~\citep{2012PhRvL.108w1101H,2012JCAP...11..058G,2021JCAP...05..034P}, thus one may wonder the limits on key features of this burst within this `loose' constraint in different scenarios of emission mechanism and the jet composition. \textbf{The neutrinos annihilation mechanism and magnetohydrodynamic processes both may affect the jet composition, which could occur in the scenarios of massive stars collapse (a single progenitor or binary progenitors, e.g. in a binary-driven hypernova model in~\cite{Aimuratov_2023}) and mergers of compact stars. }
 
The paper is organized as follows: 
in Section~\ref{sec:nuflu}, the limits for key physical parameters of GRB 230307A are extracted within the constraint of the neutrino fluence in different GRB-neutrino scenarios; in Section~\ref{sec:TRana}, basic  properties of the prompt emission for GRB 230307A are introduced and discussed; the discussion and summary are given in Section~\ref{sec:discussion}.

\section{The possible neutrino emissions from GRB 230307A }\label{sec:nuflu}
In this section, we consider the possible neutrino emissions via hadronuclear processes as well as $p\gamma$ interactions. 
For convenience, we use spectral parameters of GRB 230307A in $[2.2, 11.7]$ s after the trigger time: $\alpha=-0.8$, $\beta=-8$ and $E_{\rm peak}=1.1$ MeV. The duration is taken to be 40 s and $E_{\gamma, \rm iso}=4\times10^{52}$ erg. For comparison, GRB 221009A is also analyzed, with parameters of $\alpha=-1.0$, $\beta=-2.3$, $E_{\rm peak}=1.0$ MeV, a duration time of 327 s and $E_{\gamma,\rm iso}=2\times10^{54}$ erg. The convention $Q = 10^{n}Q_{n}$ is adopted for CGS units hereafter.

\subsection{photomeson ($p\gamma$) process}
 The decay chains of photomeson interactions for single-pion production are 
\begin{eqnarray}
\rm p+\gamma\rightarrow p +\pi^0, \label{eq:decay0}\\
\rm \pi^0\rightarrow\gamma+\gamma,
\end{eqnarray}
and 
\begin{eqnarray}
\rm p+\gamma\rightarrow n +\pi^+, \label{eq:decay1}\\
\rm \pi^+\rightarrow\mu^+ + \nu_\mu ,\label{eq:decay10}\\
\rm \mu^+\rightarrow e^+ + \nu_e +\overline{\nu}_\mu \label{eq:decay2}.
\end{eqnarray}
In the processes shown in (\ref{eq:decay1})-(\ref{eq:decay2}), one pion could produce four leptons. Before deducing formulae, some important definitions are introduced as below:
\begin{itemize}
    \item the fractions of energy dissipated in protons, random magnetic fields (with a strength $B$), and radiated as $\gamma$-rays are $\epsilon_{p}$, $\epsilon_{\rm B}$ and $\epsilon_{\rm \gamma}$.  The proton energy is estimated via $E_{\rm proton}=\epsilon_{p}/\epsilon_{e}E_{\gamma, \rm iso}$, with the CR loading factor $\xi_{\rm CR}=\epsilon_{p}/\epsilon_{e}$; note that the mark ($^{\prime}$) in the superscript denotes that the quantity is in the comoving frame, while an overline denotes the rest frame of proton, hereafter; one has $B^{\prime} = \left[\frac{L_{\rm tot} (\epsilon_B/\epsilon_e)}{2 R^2_d \Gamma^2 c}\right]^{1/2}$ in the comoving frame, where $L_{\rm tot}$ (or $E_{\rm tot}$) is the total luminosity (or energy); 
    
    \item the energy distribution of accelerated protons by random magnetic fields is $\frac{dN_{p}}{dE_{p}}\propto E_{p}^{-p}$, where the index $p=2$ is assumed; $E_{p,\rm max}$ and $E_{p,\rm min}$ represent the maximum and minimum energies of accelerated protons in the observer's rest frame; $\frac{dN_p}{dE_p} = \frac{(\epsilon_p/\epsilon_e)E_{\rm tot}}{{\rm ln}(E_{p, {\rm max}}/E_{p, {\rm min}})} E_p^{-p}$, where $E_{\rm proton}=\epsilon_{p}E_{\rm tot}$; the minimum energy of proton is set to be $E_{p, \rm min}=\Gamma m_{p}c^2$ assuming it is rest in the comoving frame; the maximum energy $E_{p, \rm max}$ is the energy of protons which are accelerated by the magnetic field, and estimated by equaling the dynamical timescale $t_{\rm dyn}^{\prime} \sim R / \Gamma c$ with the accelerating timescale of protons $t_{\rm acc}^{\prime} \sim E_p^{\prime} / (e B^{\prime} c)$;
    \item $R_{\rm d}$ is the dissipation radius: for photospheric (PH) emissions, $R_{\rm d,PH}\sim 10^{11}$ cm; for IS mechanism,  $R_{\rm d,IS}\sim 10^{12-13}$ cm; for ICMART mechanism,  $R_{\rm d,ICMART}\sim 10^{15}$ cm; they could be estimated from the parameters of the burst;
    
    \item the cross section of $p\gamma$ interaction is $\sigma_{p\gamma}=5\times 10^{-28}$ cm$^{-2}$; $\pi^{+}$ decay time scales is $\tau_{\pi^{+}}=2.8\times10^{-8}$ s; in the comoving frame, the Lorentz factor of $\pi^{+}$ is denoted as $\gamma^{\prime}_{\pi^{+}}$.  $\varepsilon_\gamma$ denotes the specific energy of a photon; $\gamma_{p}$, $\beta_{p}$ is the Lorentz Factor and velocity of a specific proton in the comoving spectrum, and so on;
    \item the width of the Breight-Weigner form of $\Delta(1232)$ resonance is $\Delta\overline{\varepsilon}_{\rm pk}\sim0.2$ GeV; $\overline{\varepsilon}_{\rm pk}\simeq0.3$ GeV are the photon energy at the resonance peak in the photomeson process.
\end{itemize}

  The pion production rate is deduced to calculate the neutrino spectrum\footnote{The followed deduction could be found in many works~\citep[e.g.][]{2022arXiv220206480K} and references therein.}.  $\sigma_{p\gamma}$ is given as a function of the photon energy in the proton-rest frame, $\overline{\varepsilon}_{\gamma}=\gamma_p^{\prime} \varepsilon_\gamma (1-\beta_p^{\prime}\mu)$, where $\gamma_p^{\prime}=\varepsilon_p^{\prime}/(m_pc^2)$, $\beta_p^{\prime}$, $\varepsilon_\gamma^{\prime}$,  $\mu=\cos\theta_p$ are the Lorentz factor of protons, the proton velocity, the photon energy, and the angle between the directions of interacting proton and photon in the comoving frame, respectively. The energy loss rate  of protons by pions production is written as
\begin{equation}
 t_{p\gamma}^{-1} = c\int d\Omega\int d\varepsilon_\gamma^{\prime} (1-\beta_p^{\prime}\mu)n_\gamma(\varepsilon_\gamma^{\prime},~\Omega)\sigma_{p\gamma}(\overline{\varepsilon}_{\gamma})\kappa_{p\gamma}(\overline{\varepsilon}_{\gamma}),
\end{equation}
where $n_\gamma(\varepsilon_\gamma^{\prime},~\Omega)=dN/(d\varepsilon^{\prime}_\gamma dV d\Omega)$ and $\kappa_{p\gamma}$ is the inelasticity of the $p\gamma$ interactions. This rate is approximately equivalent to the pion production rate. Then, to convert the integration variable from $\mu$ to $\overline{\varepsilon}_\gamma$, the photomeson production rate is represented as
\begin{equation}
 t_{p\gamma}^{-1}=\frac{c}{2\gamma_p^{\prime 2}}\int_{\varepsilon_{\rm th}}^{\infty} d\overline{\varepsilon}_\gamma\sigma_{p\gamma}\kappa_{p\gamma}\overline{\varepsilon}_\gamma \int_{\overline{\varepsilon}_\gamma/(2\gamma^{\prime}_p)}^{\infty}\frac{d\varepsilon^{\prime}_\gamma}{\varepsilon_\gamma^{\prime2}}n_{\varepsilon^{\prime}_\gamma},\label{eq:tpgam}
\end{equation}
 where we use $n_\gamma(\varepsilon^{\prime}_\gamma,~\Omega)=n_{\varepsilon^{\prime}_\gamma}/(4\pi)$, $n_{\varepsilon^{\prime}_\gamma}=dN/(d\varepsilon^{\prime}_\gamma dV)$, $\beta_p^{\prime}\sim1$. Note that the energy threshold $\varepsilon_{\rm th}$ is determined by the Breit-Weigner mass and width of $\Delta(1232)$ resonance. We have
 \begin{equation}
\sigma_{p\gamma}\kappa_{p\gamma}\approx\sigma_{\rm pk}\kappa_{\rm pk}\Delta\overline{\varepsilon}_{\rm pk}\delta(\overline{\varepsilon}_\gamma-\overline{\varepsilon}_{\rm pk}). \label{eq:resonant}
\end{equation}
The pion production efficiency, or the fraction of CR protons producing pions is given by $f_{p\gamma}={\rm min}(1,t_{p\gamma}^{-1}/t_{\rm dyn}^{-1})$.
The distribution of the target photons in the form of 
\begin{eqnarray}
n_{\gamma}(\varepsilon_\gamma) &=& \frac{dN_{\gamma}(\varepsilon_{\gamma})}{d\varepsilon_{\gamma}} \nonumber \\
&=& n_{\gamma,b}  \left\{\begin{array}{cc}
\varepsilon_{\gamma,b}^{-\alpha} \varepsilon_{\gamma}^{\alpha}, & \varepsilon_\gamma < \varepsilon_{\gamma,b}\\
\varepsilon_{\gamma,b}^{-\beta} \varepsilon_{\gamma}^{\beta}, & \varepsilon_\gamma \geq \varepsilon_{\gamma,b},
\end{array}
\right.
\label{eq:photon_spectrum}
\end{eqnarray} where 
$\varepsilon_{\rm \gamma, b}=\frac{(\alpha-\beta)E_{\rm peak}}{\alpha+2}$ and $n_{\gamma,b}$ is the specific photon number of $\varepsilon_{\rm \gamma}=\varepsilon_{\rm \gamma, b}$.
About half of the $p\gamma$ inelastic interactions produce charged pions with the energy of $\varepsilon_\pi\approx0.2\varepsilon_p$. The high-energy pions lose their energies by either synchrotron radiation or dynamical expansion before they decay to neutrinos. 
Considering the average energy fraction in pion decaying to four leptons, $\varepsilon_{\nu}=1/4\varepsilon_{\pi}$ and half decays to charged pions in $p\gamma$ inelastic interaction, the differential total muon neutrino energy is
\begin{equation}
 E_\nu^2N_{E_\nu}\approx\frac{1}{8}f_{p\gamma}f_{\pi,\rm sup}{E^2_{p}}N_{E_{p}},\label{eq:dNdEpi}
 \end{equation}
 where $f_{\pi,\rm sup}$ is the pion cooling suppression factor, and approximated to be 
 \begin{equation}
 f_{\pi,\rm sup}\approx 1-\exp(-\frac{t^{\prime}_{\pi,\rm cl}}{t^{\prime}_{\pi,\rm dec}}), \label{eq:fpicl}
 \end{equation}
 where $t^{\prime}_{\pi,\rm cl}=6\pi m_\pi^4c^3/(m_e^2\sigma_TB^{\prime2}\varepsilon^{\prime}_\pi)$ is the pion cooling time, and $t^{\prime}_{\pi,\rm dec}=\varepsilon^{\prime}_\pi \tau_\pi/(m_\pi c^2)$ is the decay time of charged pions.
 If $\varepsilon^{\prime}_{\pi}<\varepsilon^{\prime}_{\rm \pi,cooling}$ ($\varepsilon^{\prime}_{\rm \pi, cooling}=\frac{\Gamma}{1+z}\sqrt{\frac{6\pi m_\pi^5c^5}{m_e^2\sigma_T\tau_\pi B^{\prime2}}}$ is the critical cooling energy estimated by equating $t^{\prime}_{\pi,\rm syn}$ and $t^{\prime}_{\pi,\rm dec}$), the cooling effect is not efficient, and $f_{\pi,\rm sup}\approx1$, while $f_{\pi,\rm sup}\approx(\varepsilon^{\prime}_\pi/\varepsilon^{\prime}_{\pi,\rm cooling})^2$ for $\varepsilon^{\prime}_{\pi}>\varepsilon^{\prime}_{\rm \pi,cooling}$.
 
Considering the cooling process which causes the second break in neutrino spectrum\footnote{The second break is caused by $E_{p,\rm max}$ if the pion cooling is not significant.} and Equations~(\ref{eq:tpgam}$)\mbox{--}$(\ref{eq:dNdEpi}), the neutrino spectrum could be described as a power law with two breaks,
\begin{eqnarray}
n_{\nu} (E_{\nu}) &=& \frac{dN_\nu(E_{\nu})}{dE_{\nu}} \nonumber \\
&=& n_{\nu,1} \left\{ \begin{array}{cc}
\varepsilon_{\nu,1}^{\alpha_{\nu}} E_{\nu}^{-\alpha_\nu}, & E_{\nu} < \varepsilon_{\nu,1}, \\
\varepsilon_{\nu,1}^{\beta_\nu} E_{\nu}^{-\beta_\nu}, & \varepsilon_{\nu,1} \leq E_{\nu} < \varepsilon_{\nu,2},\\
\varepsilon_{\nu,1}^{\beta_\nu} \varepsilon_{\nu,2}^{\gamma_{\nu} - \beta_\nu} E_{\nu}^{-\gamma_\nu}, & E_{\nu} \geq \varepsilon_{\nu,2},
\end{array}
\right.
\label{eq:neutrino_spectrum}
\end{eqnarray}
and one has 
\begin{eqnarray}
\alpha_{\nu} = p + 1 + \beta,~\beta_\nu = p + 1 + \alpha,~\gamma_\nu = \beta_\nu + 2.
\end{eqnarray}
$\varepsilon_{\nu,1}$ could be derived in two different cases according to $\tau_{p\gamma}$. If $\tau_{p\gamma}<5$, one has
\begin{equation}\label{eq:ev10}
    \varepsilon_{\nu,1}=\varepsilon_{\nu,1}^0=6.33\times10^5 \rm GeV (1+z)^{-2} (\Gamma/300)^2 (\frac{\varepsilon_{\rm \gamma, b}}{1 MeV}) ^{-1}.
\end{equation}
if $\tau_{p\gamma}$ increases above 5,  $f_{\rm p\gamma}$ exceeds 0.5 and quickly approaches 1, and the neutrino flux no longer significantly increases with $\tau_{p\gamma}$, and $\varepsilon_{\nu, 1}$ has a form of 
\begin{eqnarray}\label{eq:ev11}
\varepsilon_{\nu, 1}=\varepsilon_{\nu, 1}^0 \times (\frac{\tau_{p \gamma}^{\rm peak}}{3})^{1+\beta},
\end{eqnarray}
where $\tau_{p \gamma}^{\rm peak}$ is the peak $p\gamma$ optical depth (the one for protons with energy $E_{p} ^{\rm peak}$ to interact with the photons with a peak energy) and described as
\begin{eqnarray}
\tau_{p \gamma}^{\rm peak} = 8.9L_{{\rm GRB},52} \left(\frac{\Gamma}{300}\right)^{-2} R_{13}^{-1} \left(\frac{\epsilon_{\gamma,b}}{\rm MeV}\right)^{-1}. 
\end{eqnarray}
$f_{\rm p\gamma\to N\pi}$ is the fraction of proton energy going into the $\pi$ production, and one has $f_{\rm p\gamma\to N\pi}=1-(1-\langle\chi_{p\to\pi}\rangle)^{\tau_{p\gamma}}$, where $\langle\chi_{p\to\pi}\rangle= 0.2$ is the the average fraction of energy transferred from protons to pions\footnote{$\tau_{p\gamma}=3$ is used in \cite{2013PhRvL.110l1101Z}, and $f_{\rm p\gamma\to N\pi}$ is much nearer to 100\% with $\tau_{p\gamma}=5$ shown in \cite{2023ApJ...944..115A}.}.

From Equation~(\ref{eq:ev10}) and (\ref{eq:ev11}), $\varepsilon_{\nu, 1}$ is denoted with
\begin{eqnarray}
 \varepsilon_{\nu, 1}=\varepsilon_{\nu, 1}^0\times \text{min}\left\{1, \left(\frac{\tau_{p \gamma}^{\rm peak}}{5}\right)^{1+\beta}\right\}.
\end{eqnarray}

$\varepsilon_{\nu,2}$ is relevant to the $\varepsilon^{\prime}_{\rm \pi, cooling}$ as 
\begin{eqnarray}
 \varepsilon_{\nu, 2}=\frac{1}{4} D\varepsilon_{\rm \pi, cooling}^{\prime}, 
\end{eqnarray}
where $D$ is the Doppler factor and $D\simeq2\Gamma$ if the observation is head on. 
The amplitude $n_{\nu,1}$ could be derived from integration of $E_{ p} \frac{dN_{p}}{dE_{p}}d E_{ p}$ from $E_{ p,\rm min}$ to $E_{ p, \rm max}$  as 
\begin{eqnarray}
 n_{\nu, 1}=\frac{1}{8} f_{\rm p}f_{\rm p\gamma\to N\pi}\frac{\epsilon_{\rm p}/\epsilon_{\rm e}E_{\gamma,\rm iso}}{\ln (\varepsilon_{\nu,2}/\varepsilon_{\nu,1})}\varepsilon_{\nu,1}^{-2},
\end{eqnarray}
where $f_{\rm p}=\frac{\ln (\varepsilon_{\nu,2}/\varepsilon_{\nu,1})}{\ln (E_{\rm p, max}/E_{\rm p, min})}$ .The observed fluence of the muon neutrinos is 
\begin{eqnarray}
 E_{\nu}^2\phi_{\nu}(E_{\nu})=\frac{E_{\nu}^2n_{\nu}(E_{\nu})}{4\pi d_{\rm L}^2}.
\end{eqnarray}
\begin{figure*}
\begin{center}
 \centering
   \includegraphics[width=.45\textwidth]{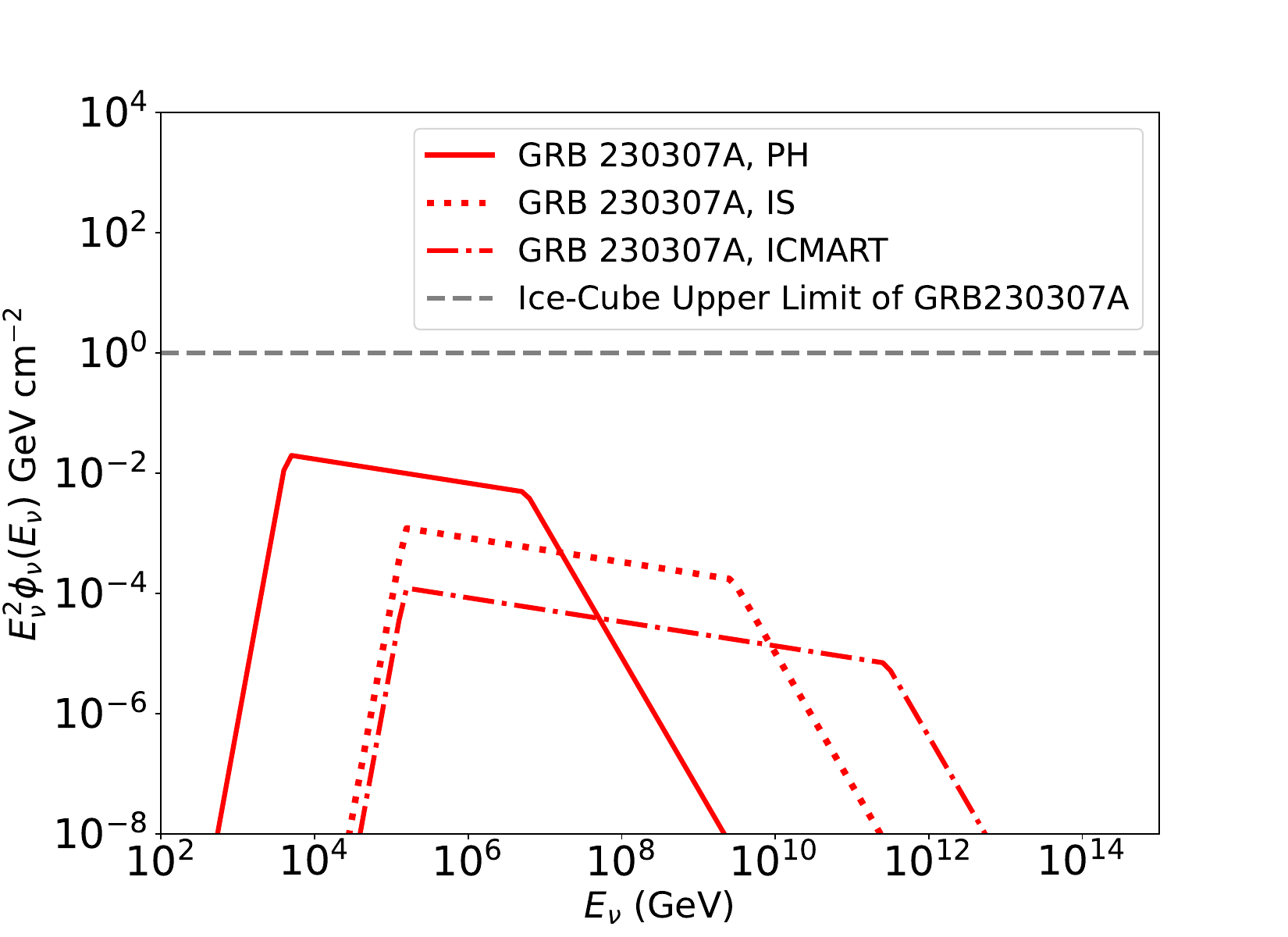}\put(-190,130){(a)}
   \includegraphics[width=0.45\textwidth]{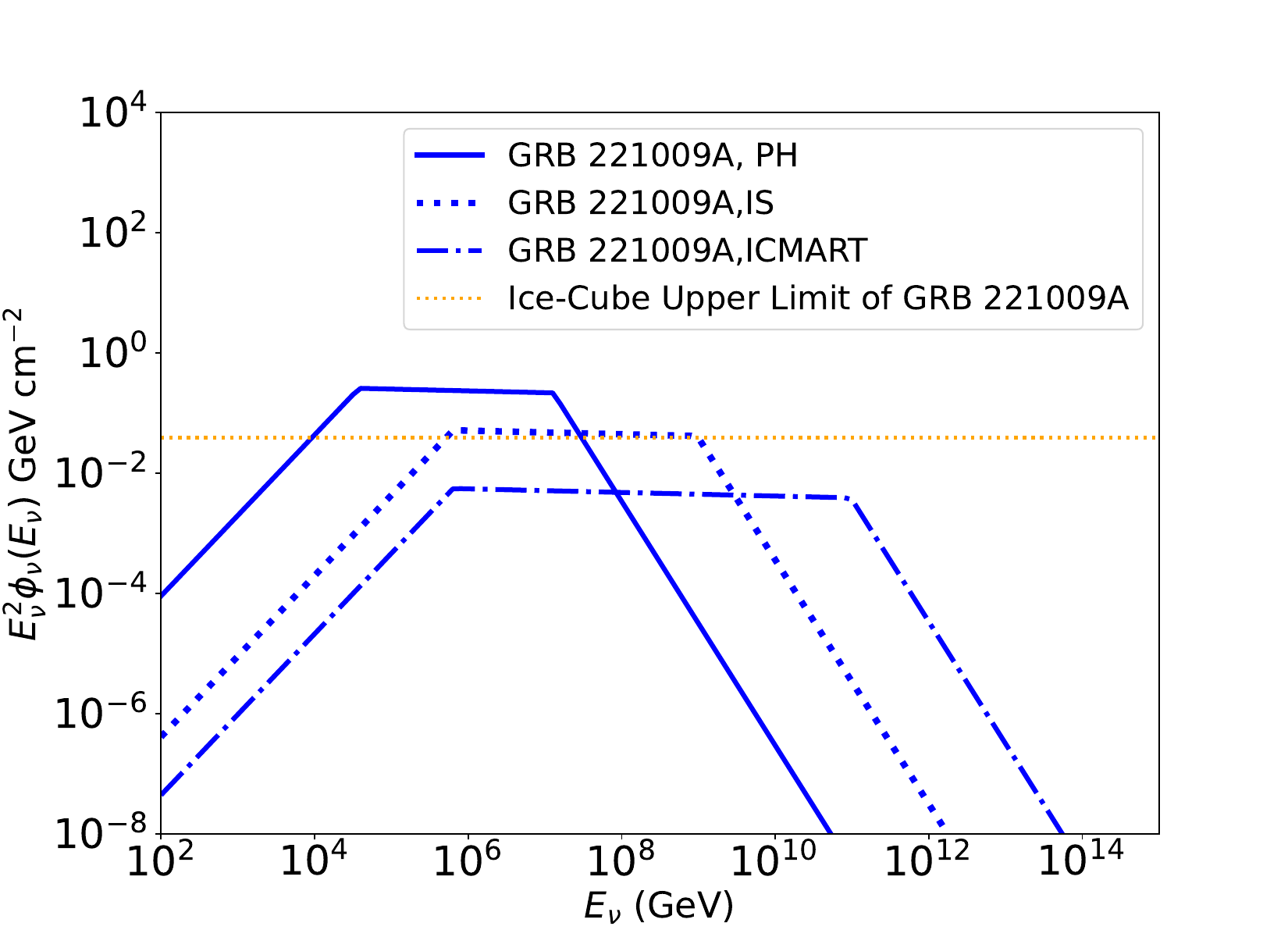}\put(-190,130){(b)}\\
  \includegraphics[width=.5\textwidth]{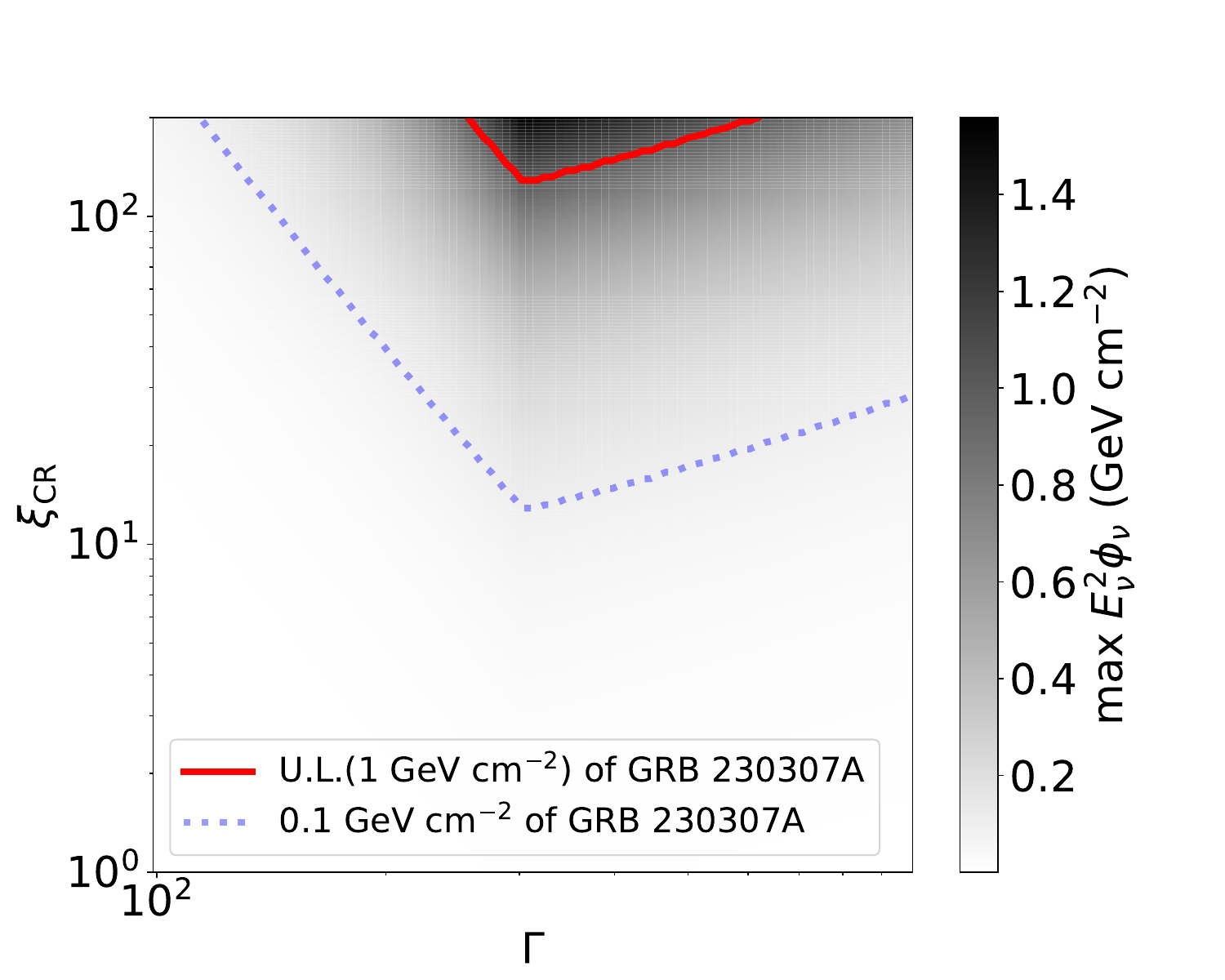}\put(-140,105){(c) $R_d=10^{11}$ cm}
   \includegraphics[width=0.5\textwidth]{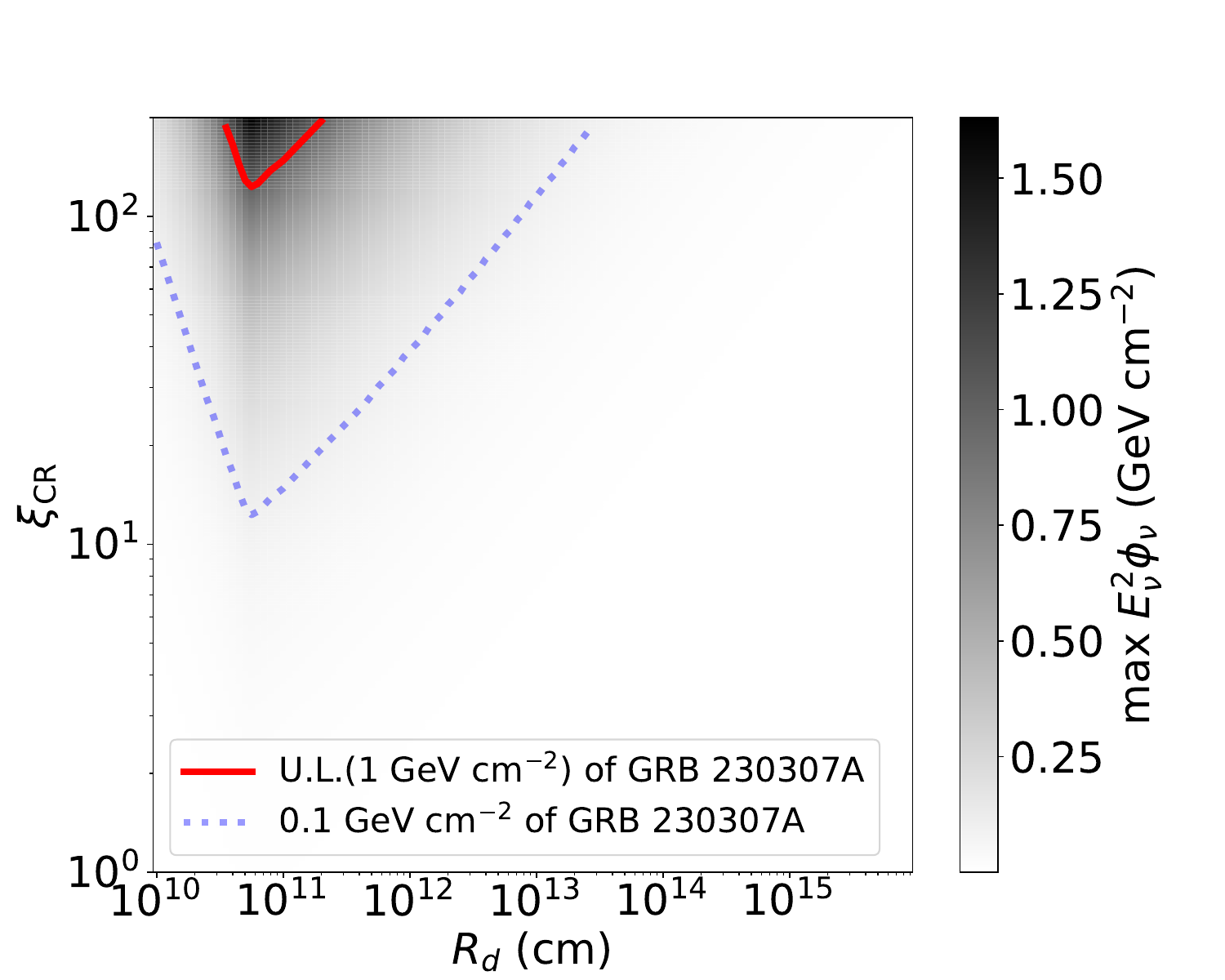}\put(-140,105){(d) $\Gamma=400$}
   \\
\caption{(a) and (b) the non-thermal neutrino spectra with $\Gamma=400$ for GRB 230307A and GRB 221009A of different emission mechanisms. For GRB 221009A, $f_{\rm p}$ is calculated to be about 0.3, 0.5 and 0.3 for PH, ICMART and IS models respectively; for GRB 230307A, $f_{\rm p}$ is calculated to be about 0.3, 0.7 and 0.5 for PH, ICMART and IS models respectively. (c) and (d) are allowed ranges of $\Gamma$, $R_d$ and $\xi_{\rm CR}$. The solid red line denotes the U.L. at 90\% C.L. of GRB 230307A. The blue dotted line denotes one order of magnitude lower than U.L. = 1 GeV cm$^{-2}$. The allowed ranges are below the corresponding lines, while the excluded ranges are above them, the same below.
\label{fig:fnu2}}
\end{center}
\end{figure*}


 With $\Gamma=400$, $\xi_{\rm CR}=\epsilon_{p}/\epsilon_\gamma=3$ and $\epsilon_{B}=\epsilon_{\gamma}$ for each emission mechanism, the predicted neutrino spectra from GRB 230307A and 221009A are shown in Figure~\ref{fig:fnu2} (a) and (b).  The fluence level of GRB 230307A with assuming a photospheric emission is the maximum of three mechanisms, which is about two orders of magnitude smaller than the U.L. of 1 GeV cm$^{-2}$, and one order of magnitude smaller than that of GRB 221009A. Furthermore, as shown in Figure~\ref{fig:fnu2} (c) and (d), $\xi_{\rm CR}$ of the excluded region is extremely large (even if with a small $R_d=10^{11}$ cm in Figure~\ref{fig:fnu2} (c)), which seems unreasonable for GRBs. The case is similar and the $\xi_N$ is still large ($>10$) if with an U.L of one order of magnitude lower (0.1 GeV cm$^{-2}$). Thus, the emission mechanisms as well as $\xi_{\rm CR}$ are not constrained within the U.L. of the neutrino production via $p\gamma$ interactions.

\subsection{hadronuclear process}
\begin{figure*}
\begin{center}
 \centering
   \includegraphics[width=.5\textwidth]{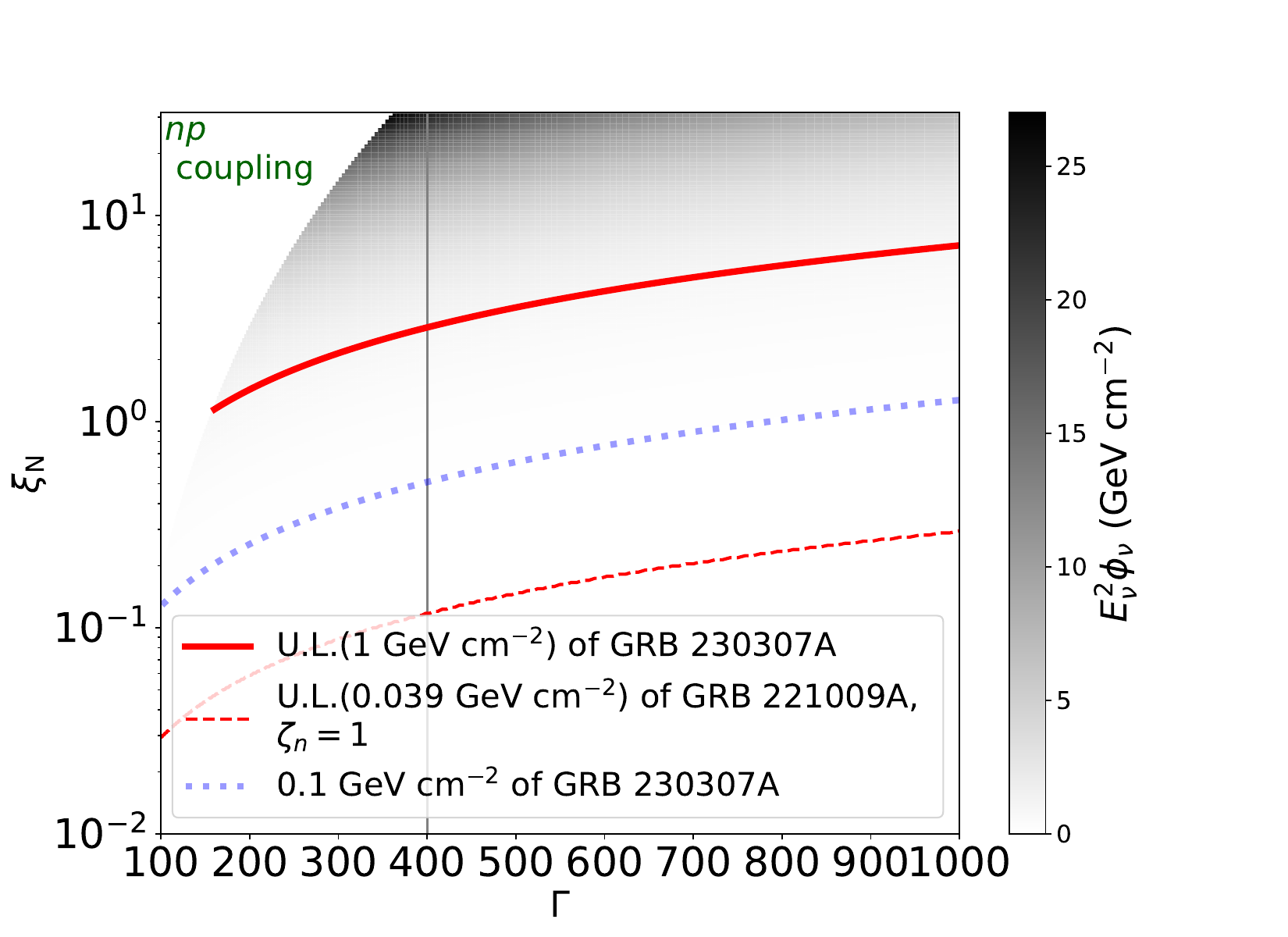}\put(-190,150){(a)$\zeta_n=5.7$ for GRB 230307A}
   \includegraphics[width=0.5\textwidth]{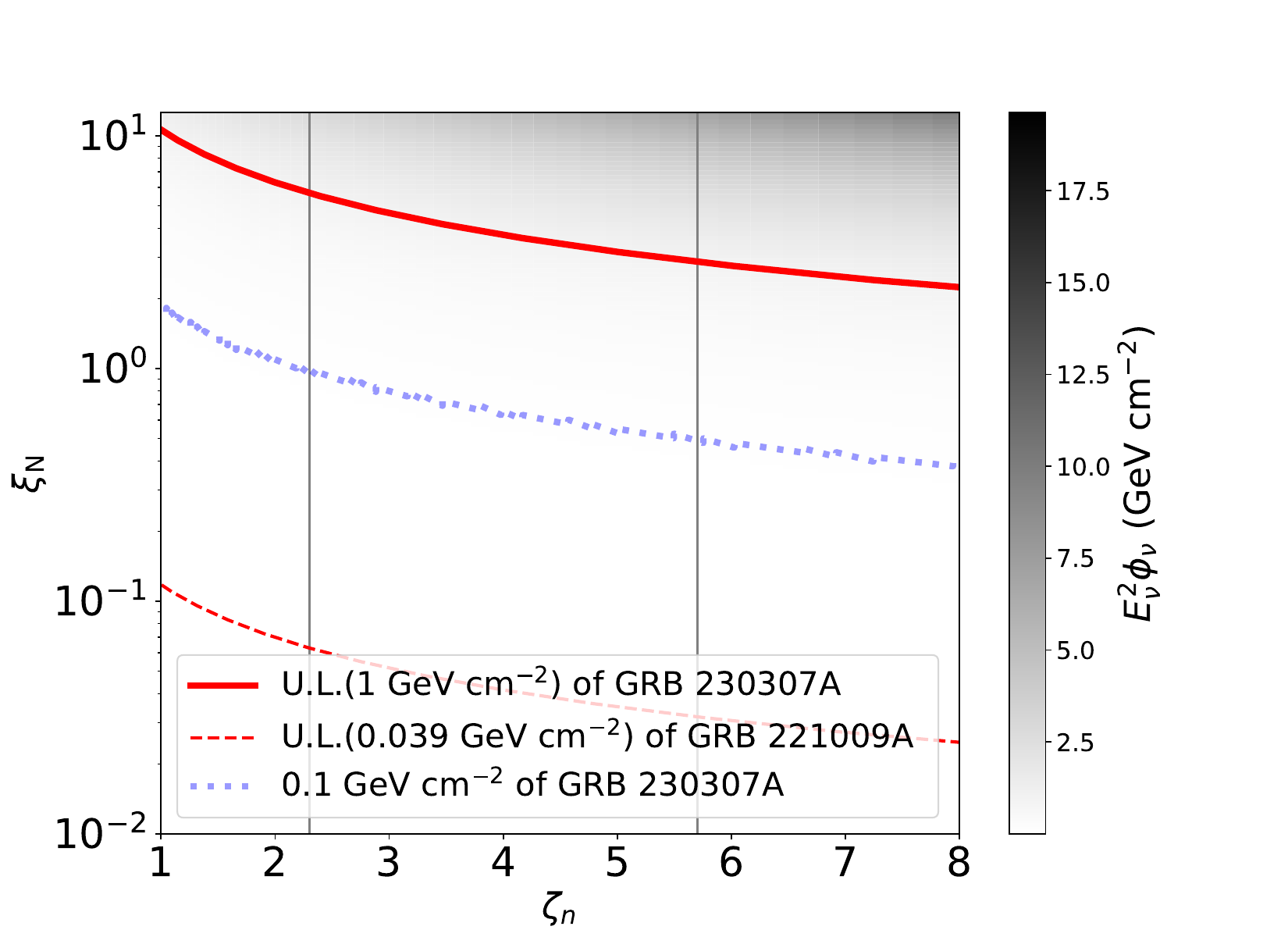}\put(-190,150){(b)$\Gamma=400$}\\
   \includegraphics[width=0.5\textwidth]{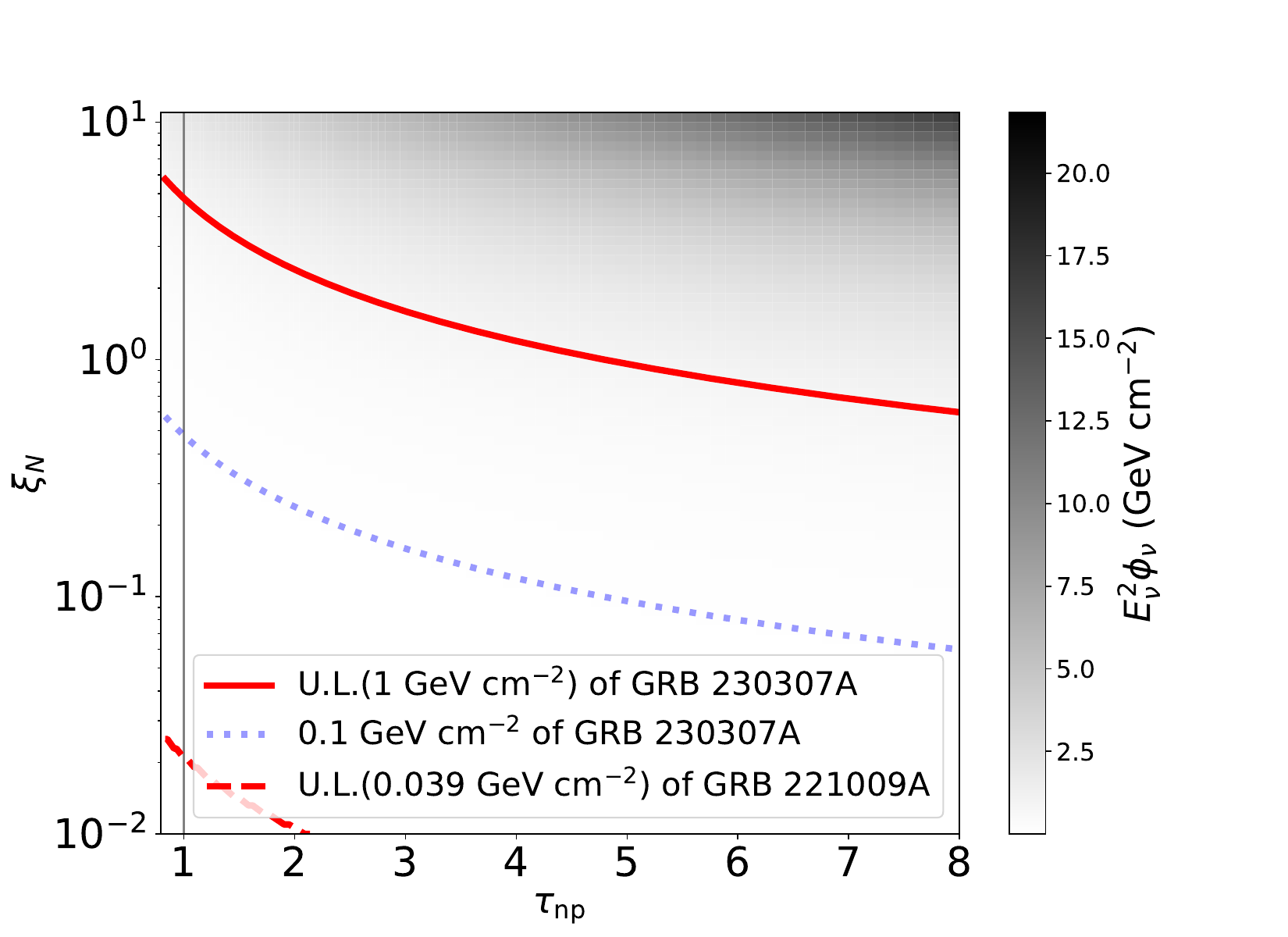}\put(-190,130){(d)} \\
\caption{(a), (b) are allowed and excluded ranges of $\Gamma$, $\xi_{\rm N}$ and $\zeta_n$ in $np$ decoupling mechanism. The blank around lower $\Gamma$ and higher $\xi_{\rm N}$ is the $np$ coupling region where $\Gamma<\Gamma_{np}$. 
The red dashed line denotes that of GRB 221009A. The gray vertical lines denote the $\Gamma=400$ for (a) and $\zeta_n=2.3$ and $5.7$ for (b).
(c) The allowed and excluded ranges of $\xi_{\rm N}$ in the $np$ collision mechanism. The vertical line denotes $\tau_{np}=1$. \label{fig:Fnu1} }
\end{center}
\end{figure*}
\subsubsection{neutron-proton ($np$) decoupling}
In the neutron-rich environment, an admixture of baryons exist in the outflow besides of photons and $e^{\pm}$. During the expanding of the outflow, the electrons can exchange energy with photons via Compton scattering and with protons via Coulomb collisions; however, for the neutrons, only protons have enough mass to collider with them and cause a collisional $np$ coupling~\citep[e.g.][]{2006MNRAS.369.1797R}. This means that, if the neutron-to-proton ratio (denoted as $\zeta_n$) is large so that there are not enough protons, or $\Gamma$ is large enough (above the the critical value $\Gamma_{np}$), the neutrons can not be accelerated so fast along with the other outflow component, even if they have drifted into the outflow in the early stage of acceleration. In this scenario the $np$ decoupling happens. After decoupling both $n$ and $p$ continue to coast together due to inertia, and their relative velocities could reach the threshold for inelastic collisions, of which decay chains are
\begin{eqnarray}
\rm p+\rm n\rightarrow \rm p + \rm n +\pi^0 , \label{eq:decay3}\\
\rm p +\rm n\rightarrow n + n + \pi^+,\label{eq:decay4}\\
\rm p + \rm n\rightarrow p+ p + \pi^-. \label{eq:decay5}
\end{eqnarray}
The charged pions and their secondary particles muons, could produce leptons via the same decay chains as those in processes~(\ref{eq:decay10})-(\ref{eq:decay2}). The detected fluence of muon neutrinos number from $np$ decoupling could be described as,
\begin{eqnarray}\label{eq:flu_qt}
\Phi_{N_{\nu_{\mu}}}=\frac{N_n N_{\nu_{\mu}}\tau_{np}}{4\pi d_L^2},
\end{eqnarray}
where $N_n$ denotes the number of neutrons contained in the outflow, $N_{\nu_{\mu}}$ denotes the number of muon neutrinos created per inelastic $np$ interaction with considering the neutrinos flavour mixing, and $\tau_{np}$ denotes the optical depth for the $np$ collision, which could be approximated to the inelastic $np$ interaction probability.

In the acceleration phase, the radius ($R$) and $\Gamma(R)$ has a relation of 
\begin{equation}
    \Gamma(R)=\Gamma_0\left(\frac{R}{r_0}\right)^{m},
\end{equation}
where $r_0$ is the initial radius of acceleration and $\Gamma_0$ is the initial Lorentz factor. $r_0$ is taken to be $10^{8}$ cm\footnote{$r_0=10^{8}$ cm is the mean value of the initial acceleration radius~\citep{2015ApJ...813..127P}.};
 $m$ ranges from 1/3 to 1. For the rapid acceleration of the fireball, $m=1$, while for the slow acceleration by the Poynting flux, $m$ could be 1/3. Note that in a collapsar scenario of a massive star, $2/3<\zeta_n\simeq1$, while in a NS-NS merger scenario, $\zeta_n\geq1$~\citep[e.g.][]{2000PhRvL..85.1362B}.  In the neutron-rich environment after the NS-NS merger for GRB 230307A, the electron (or proton) to nucleon ratio $Y_e=\frac{1}{1+\zeta_n}\sim0.15$ for a fast dynamical ejecta and $0.3$ for slow disk-wind ejecta as reported in \cite{2023GCN.33578....1B}, which corresponds to $2.3<\zeta_n<5.7$. The neutrons could penetrate the outflow, and be contained in the outflow~\citep[e.g.][]{2003ApJ...588..931B}. 

 First, the fireball scenario is considered with $m=1$. The $np$ collision time scale is $t'_{np}\approx1/(n'_p\sigma_{np}c)$, where the $\sigma_{np}\approx3\times{10}^{-26}~{\rm cm}^2$ is the approximate $np$ cross section, $n'_p\approx L_p/(4\pi R^2\Gamma\eta m_p c^3)$ is the proton density, $L_p$ is the luminosity of the proton outflow and $\eta$ is the maximum Lorentz Factor that all the energy is used to accelerate the protons. By equating $t'_{np}$ and the dynamical time scale $t'_{\rm dyn}\approx R/(\Gamma c)$, the the decoupling radius is estimated to be $r_{\rm dec}\approx8.7\times{10}^{11}~{\rm cm}~L_{p,53}^{1/3}r_{0,11}^{2/3}\Gamma_{0,1}^{-2/3}\eta_{,2.9}^{-1/3}$, at which the Lorentz factor becomes~\citep{2000PhRvL..85.1362B, 2022ApJ...941L..10M},
\begin{equation}
\Gamma^0_{n,\rm dec}\approx65~L_{p,53}^{1/3}r_{0,11}^{-1/3}\Gamma_{0,1}^{1/3}\eta_{2.9}^{-1/3}.
\label{eq:ndec}
\end{equation}
Note that $\Gamma$ must be greater than $\Gamma_{np}$ which is 
\begin{equation}
\Gamma_{np}\approx{\left(\frac{\sigma_{np}L_p\Gamma_0}{4\pi m_pc^3 r_0}\right)}^{1/4}=477 L^{1/4}_{p,53}\Gamma^{1/4}_0 r^{1/4}_{0,8}.
\end{equation}
Thus the Lorentz Factor of neutron at the decoupling time is 
\begin{equation}
\Gamma_{n,\rm dec}=\text{min}(\Gamma^0_{n,\rm dec}, \Gamma_{np})
\end{equation}
 decoupling radius by definition, and the energy of QT neutrinos is
\begin{equation}
E_\nu^{\rm QT} \approx0.1\Gamma_{n,\rm dec} m_pc^2/(1+z),
\end{equation}
which predicts $\sim1-10$~GeV neutrinos with $m_pc^2=0.938$ GeV and $\Gamma_{n,\rm dec}$ is mild ($\lesssim100$). The neutrino energy fluence is estimated to be
\begin{eqnarray}
E^2_{\nu_{\mu}} \phi_{\nu_\mu}&\approx&\frac{1}{12}\frac{(1+z)}{4 \pi d_L^2}\zeta_n\left(\frac{\Gamma_{n,\rm dec}}{\Gamma}\right)E_{\rm proton},
\end{eqnarray}
where $\phi_{\nu_\mu}=\frac{1}{4\pi d^2_L}\frac{dN_{\nu_\mu}}{dE_{\nu_\mu}}$.  
$E_{\rm proton}\approx\xi_N E_\gamma^{\rm iso}$ is the kinetic energy of the proton outflow and $\xi_N$ is the proton or nucleon loading factor. The coefficient 1/12 comes from the product of inelasticity (0.5), the fraction of charged pions (2/3) and the energy fraction of $\nu_{\mu}$ after the flavour mixing (1/4).

Figure~\ref{fig:Fnu1} (a) and (b) shows the allowed ranges of $\xi_{\rm N}$, $\Gamma$ and $\zeta_n$ within the constraint from U.L. of neutrino fluence. With $\Gamma=400$ and $\zeta_n=5.7$, the U.L. of $\xi_{\rm N}$ could reach up to 3. From Figure~\ref{fig:Fnu1} (b), the U.L. of $\xi_n$ decreases with the $\zeta_n$. 
 With one order of magnitude lower than 1 GeV cm$^{-2}$, the U.L. of $\xi_N\lesssim1$ as shown in dotted lines in Figure~\ref{fig:Fnu1} (a) and (b). Compared with the largest allowed $\xi_{\rm N}\lesssim0.1$ for GRB 221009A (with $\zeta_n=1$), it seems that a heavier baryon loading could be allowed for GRB 230307A.  

If another case that the Poynting flux is dominated ($\sigma_0>10$, where $\sigma_0$ is the Poynting-to-kinetic flux
ratio) is considered, the neutrino fluence could be smaller. This is because $\tau_{np}$ in Equation~(\ref{eq:flu_qt}) has a form of
\begin{eqnarray}
\tau_{np}(m)=\int^{\infty}_{\chi_\pi}(1-\frac{0.75}{\ln \chi}) (\chi^{-1}-\chi^{-3})\chi^{-1/m},
\end{eqnarray}
 where $\chi$ is the ratio of Lorentz Factor of proton to that of neutron and $\chi_\pi\simeq2.15$ is the threshold for pion production~\cite[e.g.][]{1971ctf..book.....L,2007A&A...471..395K}. For the Poynting-flux-dominated (PFD) case, $m=1/3$ and the saturation radius is much larger than the pion creation radius, $\tau_{np}\simeq0.01\zeta_n^{1/2}$ is several times smaller that that of fireball ($\tau_{np}\simeq0.05\mbox{--}0.1$). Thus, the corresponding U.L. of $\xi_N$ is even much larger than that obtained in the fireball scenario, which seems too large for the assumption that the Poynting flux is dominated. Therefore, the U.L. can not provide a stringent constraint in the PFD scenario.

A third scenario of a hybrid jet with $1\lesssim\sigma_0<10$, could be considered as well. If the decoupling occurs during the rapid acceleration, the case is similar to the fireball scenario; similar to the above discussion, the U.L. of $\xi_N\sim$ a few could be also obtained, which seems consistent with the hybrid scenario if we assume there exists a Poynting flux of which the energy is comparable with $E_{\gamma, \rm iso}$. Otherwise, it is similar to the PFD scenario.



\subsubsection{collisions in neutron-loaded flows}
The $pn$ collisions occur not only in the case of $np$ decoupling before the coasting regime, but also in the case of internal collisions between the compound flow if $\Gamma_{np}$ is larger than $\eta$ and the coasting occurs earlier than the decoupling~\citep[e.g.][]{2013PhRvL.111m1102M}. 
As discussed in \cite{2013PhRvL.111m1102M}, the energy of the produced neutrinos is
\begin{equation}
E_\nu^{\rm qt} \approx0.1\Gamma\Gamma_{\rm rel}^\prime m_pc^2/(1+z),
\end{equation}
where $\Gamma_{\rm rel}^\prime\sim2$ is the relative Lorentz factor of the interacting flow and sub-TeV neutrinos are expected for $\Gamma\sim{10}^2\mbox{--}{10}^3$. The neutrino energy fluence is
\begin{eqnarray}
E_\nu^2 \phi_{\nu_\mu}&\approx&\frac{1}{12}\frac{(1+z)}{4 \pi d_L^2}\tau_{pn}\xi_N{E}_{\gamma, \rm iso}.
\end{eqnarray}
 Internal collisions between neutron-loaded flows may be relevant for sub-photospheric dissipation~\citep[e.g.][]{2010MNRAS.407.1033B}.  $\xi_{N}\sim4-20$ is predicted in this scenario~\cite[e.g.][]{2011ApJ...738...77V}. As shown in Figure~\ref{fig:Fnu1} (c), the U.L. of $\xi_{\rm N}$ could reach up to $\sim5$ with $\tau_{np}=1$, which could be consistent with the scenario. The U.L. of $\xi_N\sim0.5$ is obtained within the smaller constraint of U.L=0.1 GeV cm$^{-2}$, seems too small. However, it may be reasonable for a hybrid jet in which there exists an appreciable Poynting flux, because a considerable part of $E_{\gamma, \rm iso}$ could be contributed by the dissipation of the Poynting flux via e.g. ICMART mechanism.

\subsection{Other processes: $pp$ collisions, multi-pions and Kaons production in $pp$ collisions and $p\gamma$ interactions }

Other processes, such as $pp$ collisions, multi-pions and kaons production in $pp$ collisions and $p\gamma$ interactions~\cite[e.g.][]{2009ApJ...691L..67W,2012JCAP...11..058G} could also contribute to the neutrino emission. However, compared with the estimations above, their contributions are not much larger, and do not affect the conclusion. 

\subsection{Implications}
 The U.L. (1 GeV cm$^{2}$) of the neutrino fluence for GRB 230307A seems too loose to constrain a specific GRB emission mechanism, if the non-thermal neutrino production via $p\gamma$ interactions is the only consideration. For QT neutrinos, the U.L. of $\xi_N\sim$ a few, which seems higher than that ($\xi_N\lesssim0.1$) of GRB 221009A. If the baryon loading is really appreciable, the jet is not PFD as that in GRB 221009A. Therefore, it may be necessary to directly investigate the prompt emission mechanism from spectra in the prompt phase to see if there is something different.

\section{Basic spectral properties of GRB 230307A}\label{sec:TRana}
\begin{figure*}
\begin{center}
 \centering
   \includegraphics[width=.90\textwidth]{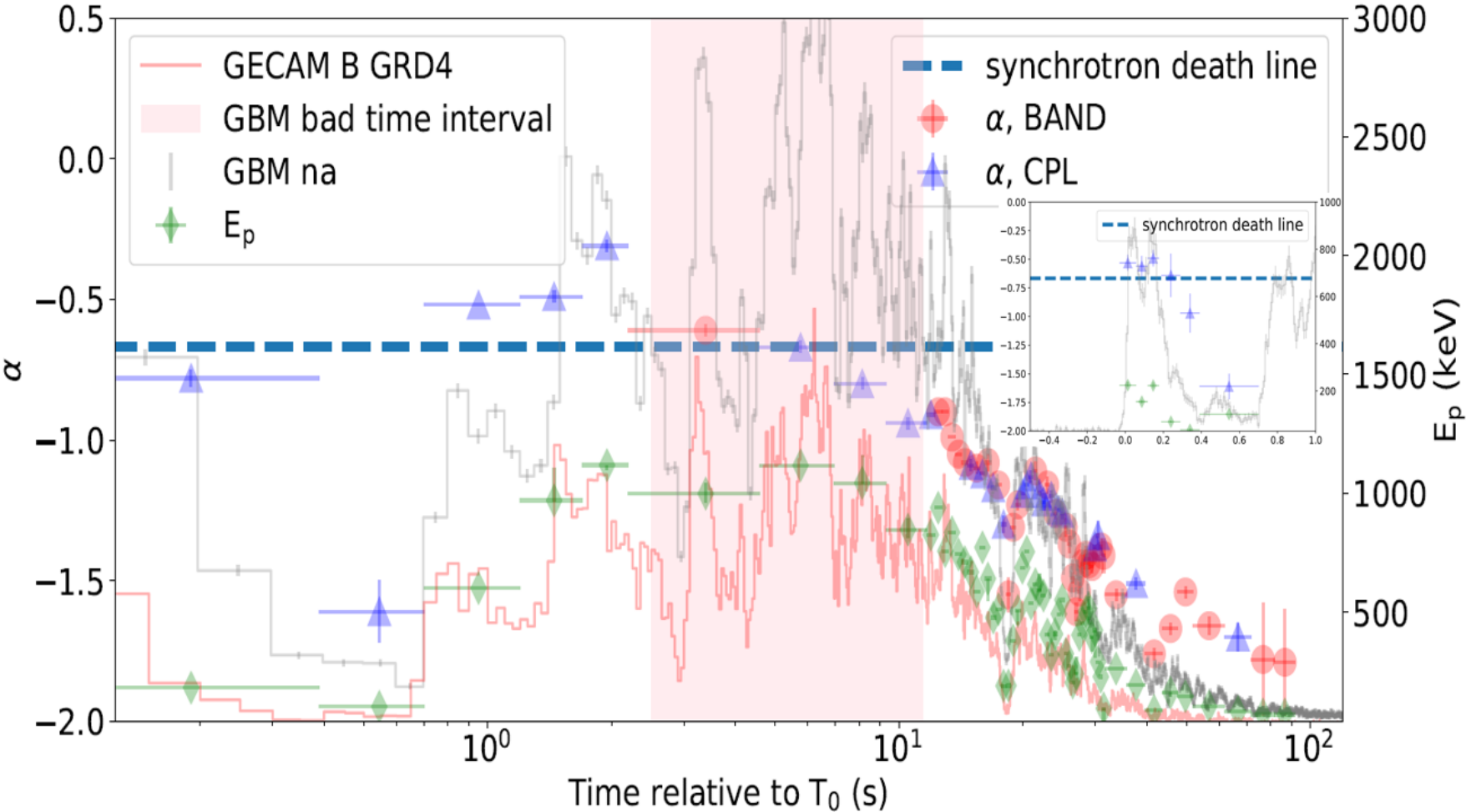}\put(-310,185){(a)} \put(-130,180){(b)}\\
  \includegraphics[width=0.33\textwidth]{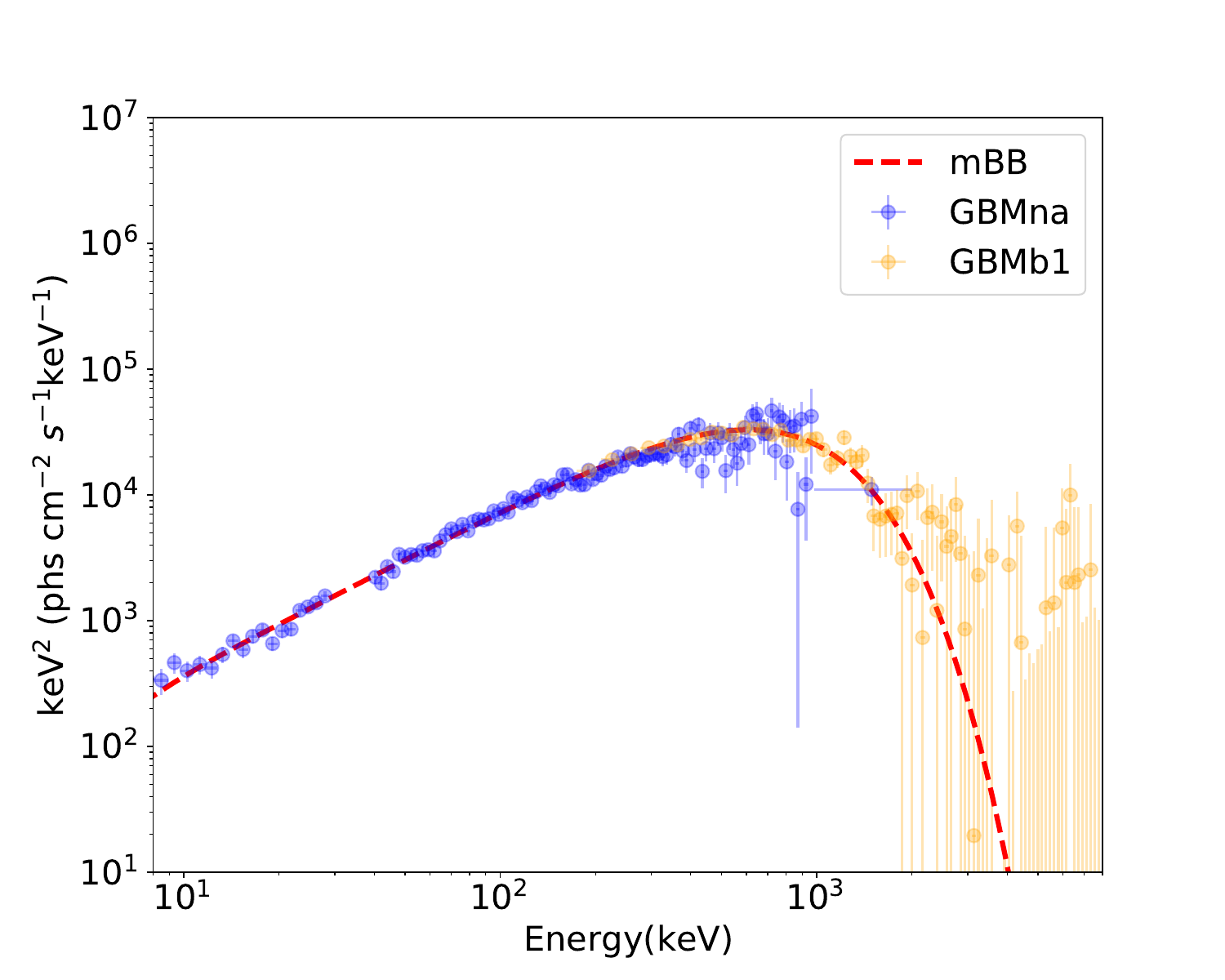}\put(-120,110){(c)[-0.7,1.2] s }
   \includegraphics[width=0.33\textwidth]{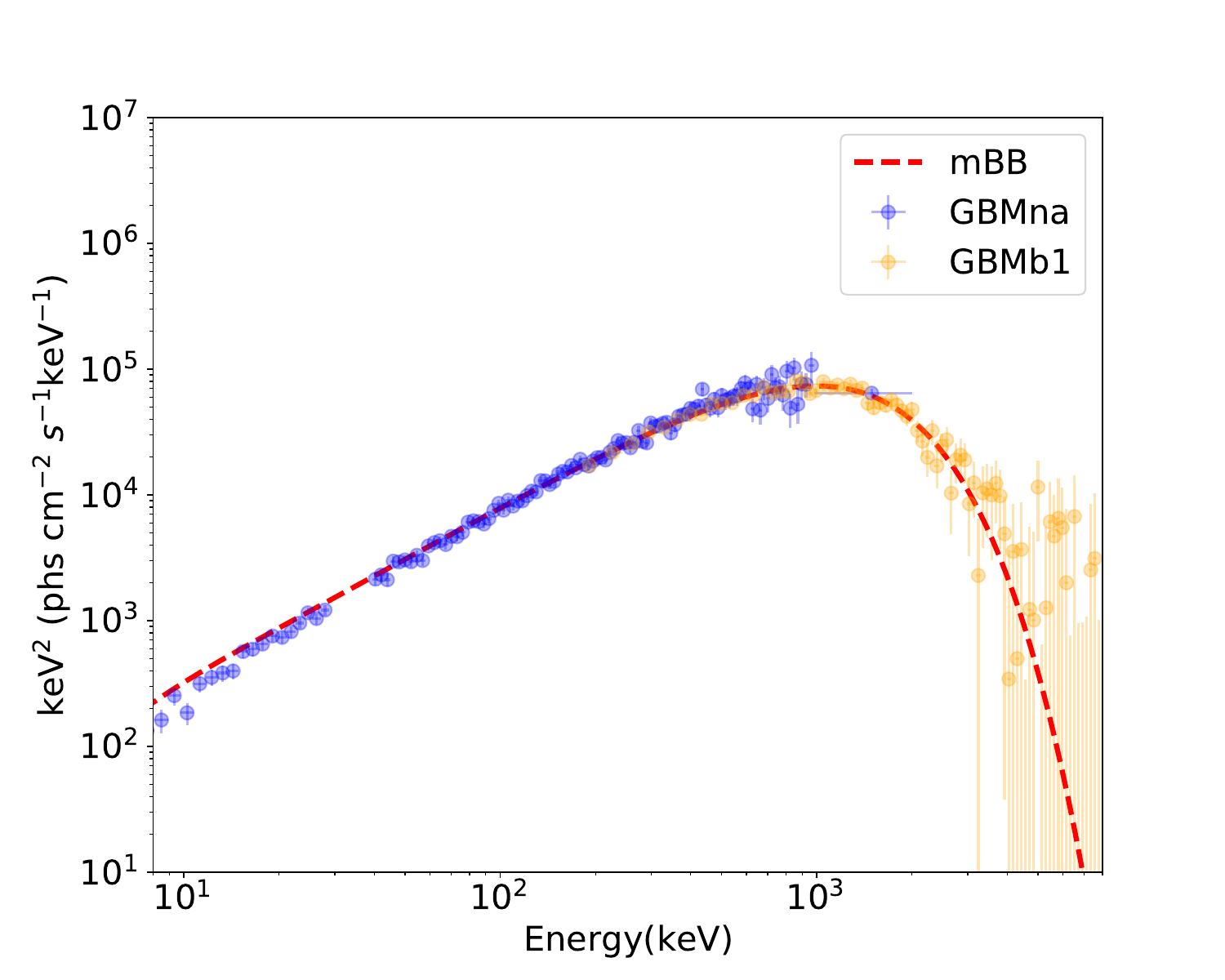}\put(-120,110){(d)[-1.2,1.7] s }
   \includegraphics[width=0.33\textwidth]{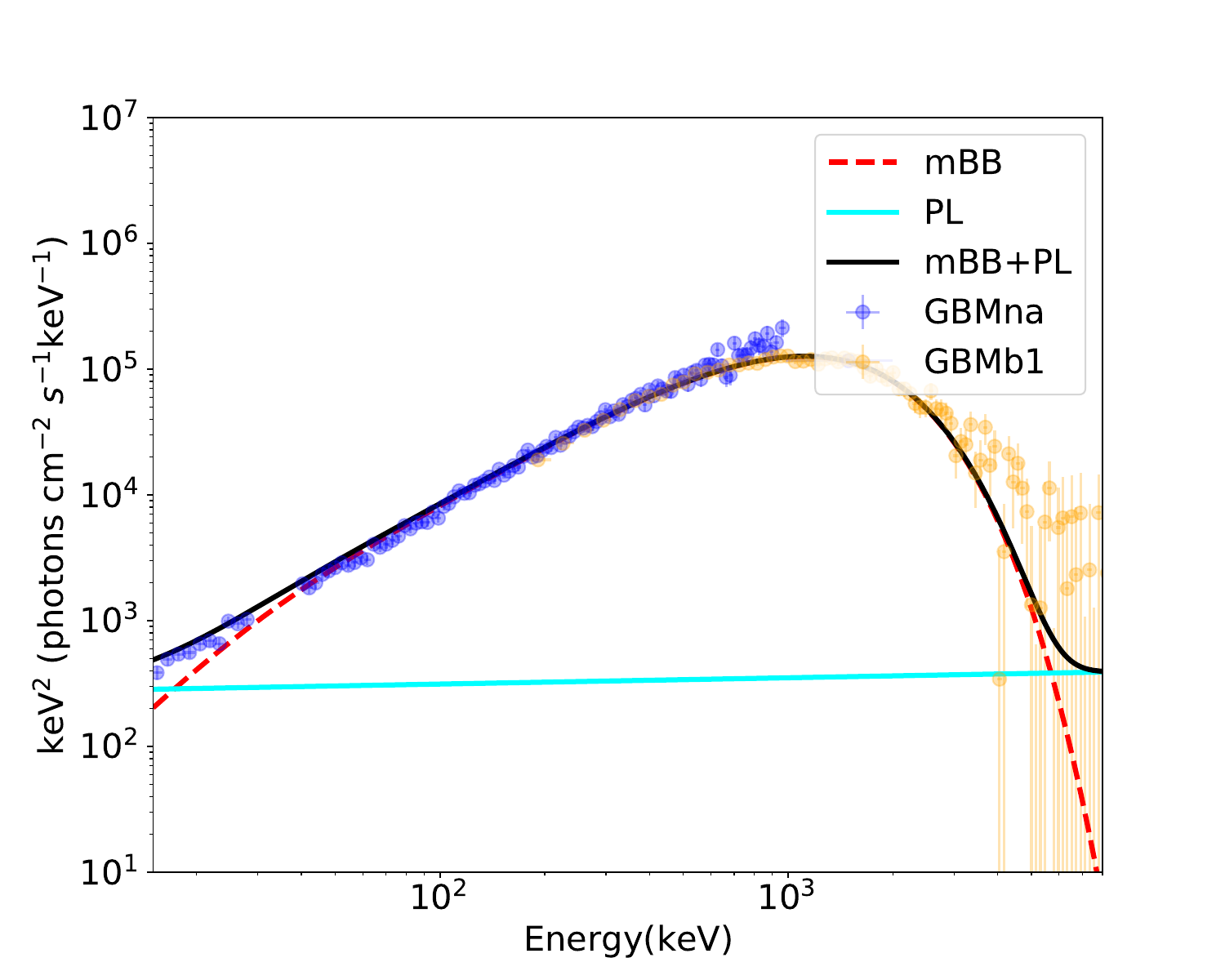}\put(-120,110){(e)[-1.7,2.2] s }
   \\
    \includegraphics[width=0.33\textwidth]{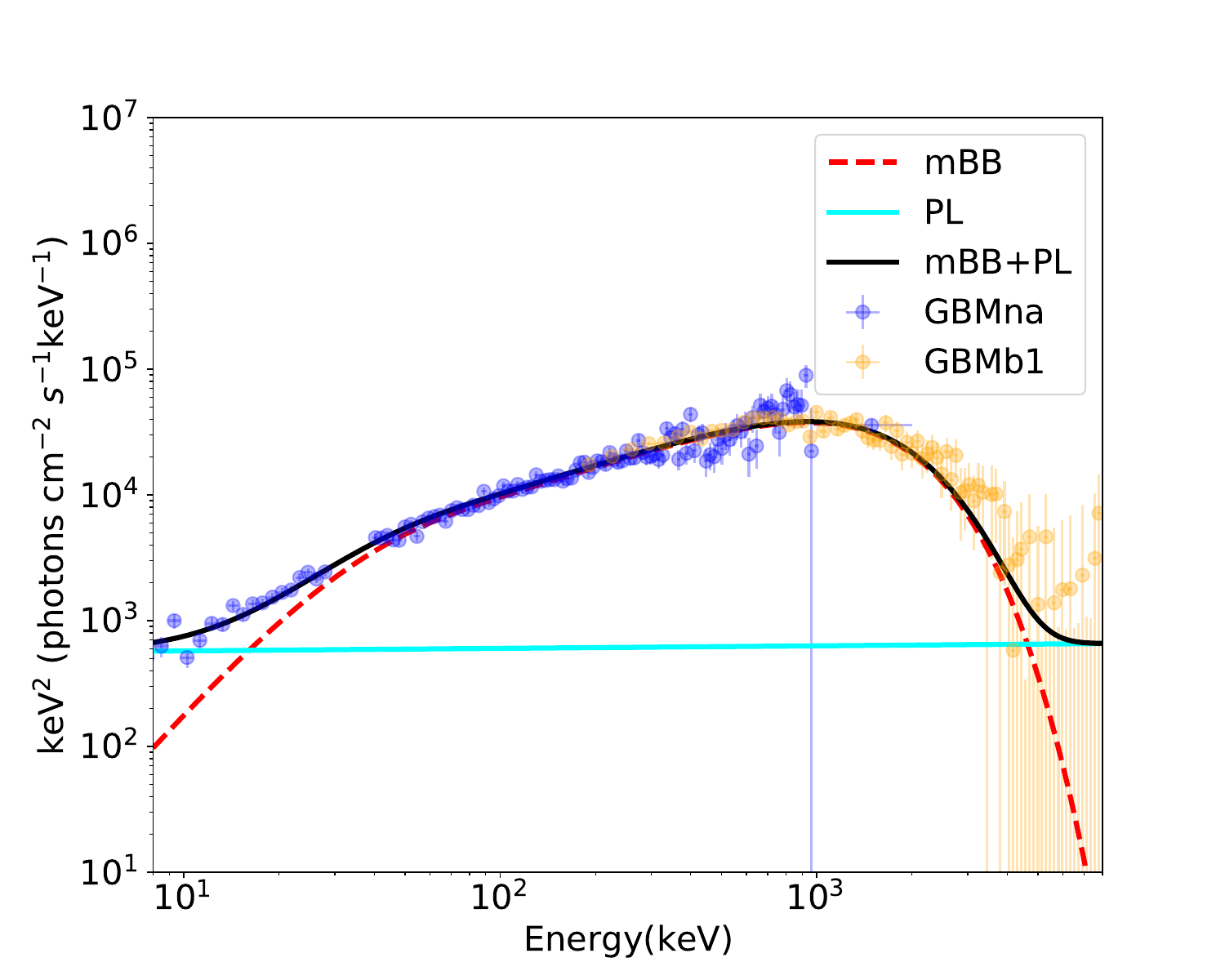}\put(-120,110){(f)[11.7,12.2] s }
   \includegraphics[width=0.33\textwidth]{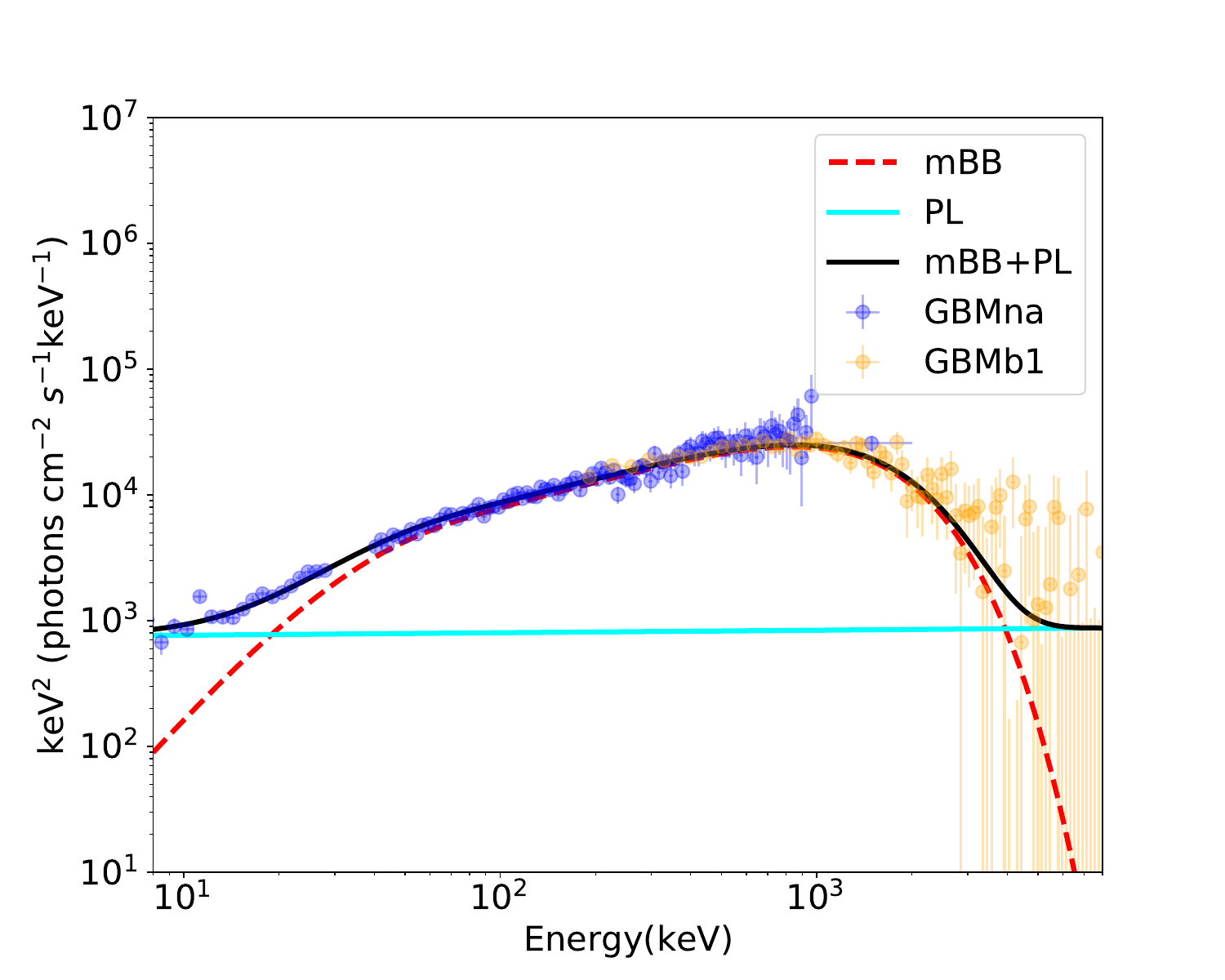}\put(-120,110){(g)[13.7, 14.2] s }
   \includegraphics[width=0.33\textwidth]{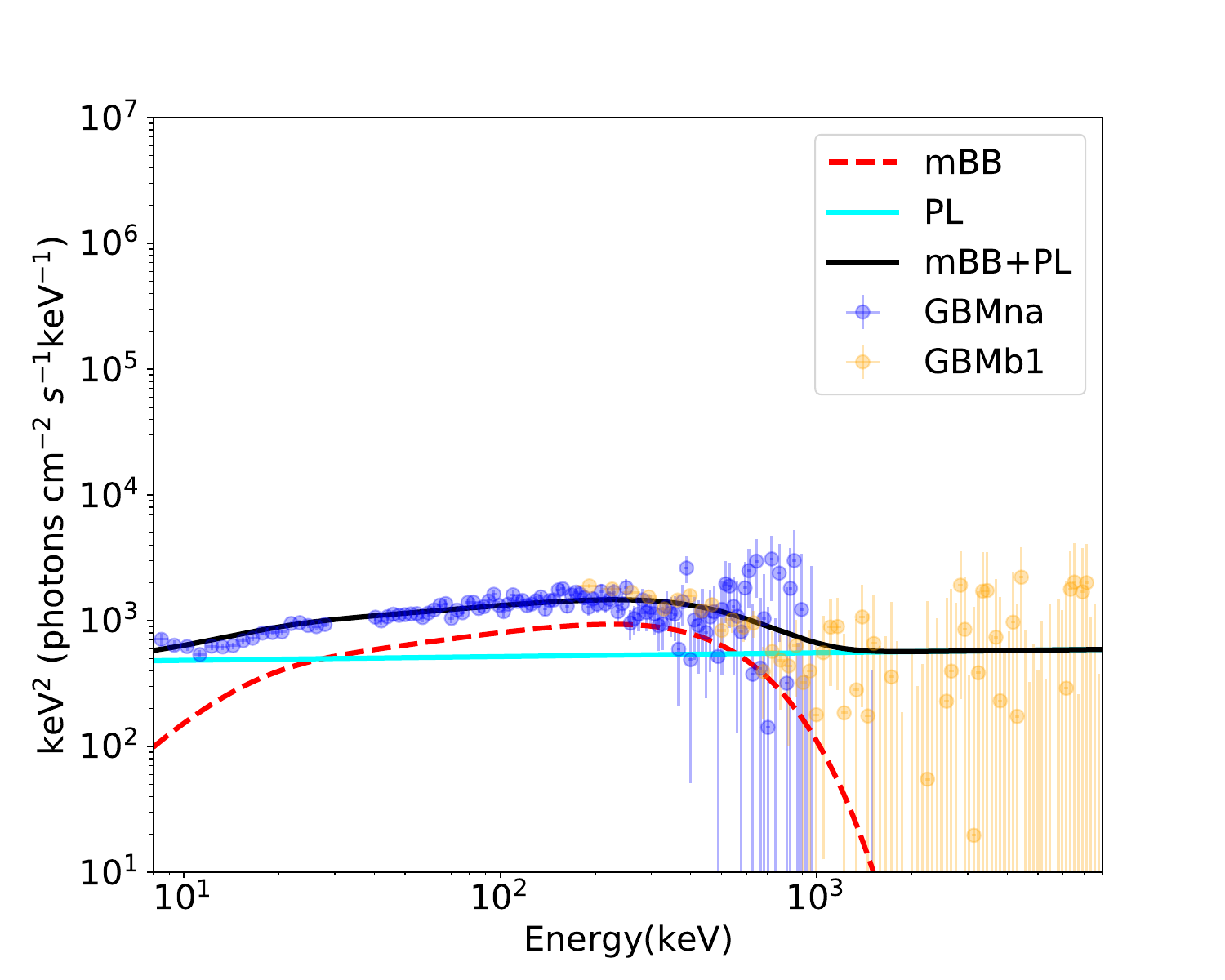}\put(-120,110){(h)[35.7, 39.7] s }

\caption{(a) The light curves (logarithm scale) of GBM na detector and $\alpha$, $E_{\rm p}$ values, with a sub-plot (b) for the first 1 s. Red circles and blue triangles denote $\alpha$ values from the fit results with BAND and CPL functions (the same below), respectively. Green diamonds denote the $E_{\rm p}$ values from the fit results. The sub-plot shows $\alpha$ and $E_{\rm p}$ values of the first 1 s. (c)-(h) Time resolved spectra with modeling with multi-blackbody (mBB) or mBB+PL functions.
\label{fig:LC}}
\end{center}
\end{figure*}

GBM NaI detector (NaI 10), and  GBM BGO (BGO 1) detector recorded the brightest flux, however, dead time (DT) and pile-up instrumental (PUI) effects occur in [$T_0+2.5$, $T_0+11.0$] s~\citep[GCN 33551,][]{2023GCN.33551....1D}, as shown in the red shadow in the light curve in Figure~\ref{fig:LC} (a). Fortunately, data from GECAM-B are not distorted by PUI effect. Therefore, data analysis is performed with data from the brightest detector (GRD 4) of GECAM-B from [$T_0+2.2$, $T_0+11.7$] s, while in other time bins, data from the two brightest detectors (NaI 10 and BGO 1) of GBM are used. A polynomial is applied to fit all the energy channels and then interpolated into the signal interval to yield the background photon count estimate for GRB data. The Markov Chain Monte Carlo (MCMC) fitting is performed to find the parameters with the maximum Poisson likelihood. 

 From Figure~\ref{fig:LC} (a), the values of $\alpha\gtrsim-2/3$~\citep[synchrotron death line,][]{Preece_1998} concentrate in the first 7 s, which implies that the emission at the beginning may have a photospheric origin. The fit results are shown in Table~\ref{tab:fitresTR} in Appendix~\ref{sec:fitrestable}.
After about 7 s, $\alpha$ gradually decreases to about -1.6, which means that the non-thermal emission may become dominant gradually afterwards.
The typical QT spectra in the first few seconds are shown in Figure~\ref{fig:LC} (c)-(e), which is narrow with the maximum of $\alpha\sim-0.3$ and more like a broadened Planck spectrum after subtracting the non-thermal emission denoted by a power-law (PL) function. The photospheric emission is a natural consequence of the fireball~\cite[e.g.,][]{1978MNRAS.183..359C,1986Are,1986ApJ308L43P}. The Planck spectrum related to the photospheric emission could be broadened by geometrical broadening~\citep{2008ApJ...682..463P,2013MNRAS.428.2430L}, as well as
by dissipations below the photosphere~\citep[e.g.,][]{2005Dissipative}, such as magnetic reconnection~\citep[e.g.,][]{2004Spectra}, and hadronic collision shocks~\citep[e.g.,][]{2010Radiative}.

From Figure~\ref{fig:LC} (f)-(h) and the decreasing $\alpha$ with time, it is found that the so-called QT spectral shape becomes broader.
This may be caused by the enhanced dissipation below the photosphere with time. The details of the spectral evolution are not the main goal in this letter and will not be further discussed. In summary, an significant existence of the photospheric emission is observed in the prompt phase of GRB 230307A, which is greatly different from that in GRB 221009A.

 \section{Discussion and Summary}\label{sec:discussion}
 \textit{I. Constraints from the possible neutrino emission on GRB 230307A} 
 
 $E_{\rm iso, \gamma}$ for GRB 230307A is not large enough, compared with the nearby extremely bright GRB 130427A and 221009A. The U.L. seems so loose that the emission mechanisms for this burst are not well constrained, with assuming that the neutrinos production from $p\gamma$ interactions is the only consideration. The case is similar even if the U.L. is one order smaller. It is proposed that the QT neutrinos from hadronuclear processes may contribute to the neutrino emission. The U.L. of allowed nucleon loading factor is about a few in this neutron-rich post-merger environment of GRB 230307A.  
 
 \textit{II. The nucleon loading and emission mechanism in GRB 230307A}
 
 The neutrino emission is relevant to the emission mechanism and the jet composition for GRBs. Three different jet properties are considered in the estimation for the QT neutrino fluence. The U.L. of the nucleon loading $\xi_{\rm N}\sim$ a few for GRB 230307A, while the  U.L. of $\xi_{\rm N}\lesssim0.1$ for GRB 221009A, which may imply that a heavier nucleon loading is allowed in GRB 230307A than that in GRB 221009A. 
 
 For GRB 221009A, the U.L. of allowed $\xi_N\lesssim0.1$ is very small, which is consistent very well with its jet composition and emission properties inferred from the spectral modeling~\citep{2023ApJ...947L..11Y}: a PFD jet (with a synchrotron-dominated prompt emission).  The prompt emission of GRB 230307A seems quite different from that of GRB 221009A. A significant photospheric emission dominates the beginning of the prompt phase; the spectrum becomes broader and broader with time, which could be interpreted by enhanced sub-photosphere dissipations via magnetic reconnection, or hadronic collisions. The dissipation via hadronic collisions could occur in a fireball or a hybrid jet, while the magnetic dissipation may happen in a hybrid or PFD jet.
 If the jet is PFD, it is inferred that there must be a thermalization via the magnetic dissipation below the photosphere, to interpret the observed photospheric emission.

GRB 211211A is another nearby GRB ($z=0.076$) and has a same origin as GRB 230307A~\citep{2022Natur.612..232Y}. There is no report about the relevant neutrino emission, however, $\sigma_0$ tells some information about the nucleon loading.  $\sigma_0\sim 10\mbox{--}100$ is estimated for GRB 211211A~\citep{2023ApJ...943..146C}. This means that the jet for GRB 211211A is PFD and nucleon loading is small. Besides, there exist neither significant photospheric emissions, nor evident spectral evolution (from photospheric to non-thermal) in the prompt phase. It may be inferred that the jet composition of GRB 230307A is somewhat different from those of GRB 211211A and GRB 221009A, and a heavier baryon loading may account for it.

 \textit{III. Assuming a more stringent constraint from neutrinos detection }
 
 A data set of high-energy track events are used in the analysis by IceCube to extract the U.L. of neutrino fluence, and the sensitivity can vary greatly over different declinations ($\delta$) ~\citep{2021ApJ...910....4A,2023ApJ...946L..26A}. GRB 230307A is located at the southern hemisphere with $\delta=-75.4^\circ$, while  $\delta=+19.8^\circ$ for GRB 221009A. The U.L. for non-detection in GRB 230307A might be about one order of magnitude lower if it was in northern hemisphere with a reduced background. Thus, with assuming that it has the same sensitivity as that in the northern hemisphere, 0.1 GeV cm$^{-2}$ as a U.L. is also used in the analysis as shown in Section~\ref{sec:nuflu}. However, this smaller U.L. still can not provide a stringent constraint on GRB emission mechanisms via $p\gamma$ interactions, as shown in  Figure~\ref{fig:Fnu1}. For the QT neutrino emission, a smaller U.L. is obtained, however, it is still much larger than that in GRB 221009A or GRB 211211A. The conclusion is similar to that with 1 GeV cm$^{-2}$ as a U.L..
 
 \textbf{We note that till now no GRB-neutrinos have been ever detected, even for the two brightest nearby GRBs ever observed (GRB 221009A and GRB 230307A), which have different representative dissipation mechanisms. There may be some hints that: only a brighter and/or more nearby GRB might be expected for the detection for the candidate of the source of GRB-neutrinos;
  for the future neutrino detectors, sub-TeV channels and relevant dedicated searches with considering the neutrino spectrum shape are of great importance. }


\begin{acknowledgements}
 The authors thank supports from the National Program on Key Research and Development Project (2021YFA0718500) and National Natural Science Foundation of China (grant Nos. 12303052). This work is partially supported by International Partnership Program of Chinese Academy of Sciences (Grant No.113111KYSB20190020). The author is very grateful to the suggestions from Jessie Thwaites and the IceCube Collaboration. Dr. Xin-Ying Song thanks public GRB data of Fermi/GBM and GECAM-B data group. I am very grateful for the comments and suggestions from the anonymous referees. I thank Dr. Rui Qiao for his suggestions on the response matrix generating for GECAM-B data. I thank Dr. Michael S. Briggers and Dr. Michelle Hui for their suggestions on GBM data. I thank Dr. Xi-Lu Wang and Prof. Shuang-Nan Zhang for their suggestions.
\end{acknowledgements}
\appendix
\section{The fit results of the prompt emission of GRB 230307A}\label{sec:fitrestable}
The fit results modeling with mBB or mBB+PL models and empirical functions are listed in Table.~\ref{tab:fitresTR}. 
\clearpage
\renewcommand\thetable{\Alph{section}\arabic{table}}  
\begin{deluxetable*}{lccccccccc}
\tabletypesize{\tiny}
\tablewidth{0pt}
\tablecaption{The time-resolved results with mBB, mBB+PL and BAND functions.\label{tab:fitresTR}}
\tablehead{
\colhead{Time bins}
& Favoured or Used Model
&\colhead{$m$}
&\colhead{$kT_{\rm min}$}
&\colhead{$kT_{\rm max}$}
&\colhead{$F_{\rm T}$}
&\colhead{$\alpha$}
&\colhead{$F_{\rm NT}$}
&\colhead{BIC}
&\colhead{$\frac{\chi^2}{ndof}$}\\
\colhead{(s)}
&\colhead{}
&\colhead{}
&\colhead{(keV)}
&\colhead{(keV)}
&\colhead{(10$^{-6}$ erg cm$^{-2}$ s$^{-1}$)}
&\colhead{}
&\colhead{(10$^{-6}$ erg cm$^{-2}$ s$^{-1}$)}
&\colhead{}
&\colhead{}}
\startdata
[-0.01, 0.39] & &-0.35$^{+0.07}_{-0.09}$ &5.9$^{+0.7}_{-0.4}$ &105.2$^{+7.0}_{-5.2}$ &30.3$^{+0.5}_{-0.4}$ &  &  &249.0 &$\frac{239.7}{207}$\\
&mBB+PL&-0.44$^{+0.09}_{-0.09}$ &8.5$^{+1.2}_{-1.2}$ &108.5$^{+50.6}_{-20.0}$ &27.2$^{+1.8}_{-1.2}$ &-2.09$^{+0.05}_{-0.01}$ &7.0$^{+7.2}_{-2.5}$ &235.6 &$\frac{221.6}{205}$\\\hline
[0.39, 0.7] &BAND &$\alpha$ &$\beta$ &$E_{\rm p}$ &  &  &  & & \\
& &$-1.61^{+0.24}_{-0.12}$ & $-6.88^{+3.66}_{-3.49}$ &$101.3^{+17.0}_{-12.7}$ &  &  &$5.89^{+0.89}_{-0.61}$  &122.3 &$\frac{113.6}{147}$\\\hline
[0.7, 1.2] &mBB &0.30$^{+0.02}_{-0.03}$ &1.9$^{+1.0}_{-0.6}$ &259.9$^{+5.2}_{-4.6}$ &106.6$^{+1.1}_{-1.5}$ &  &  &250.2 &$\frac{240.6}{236}$\\
& &0.24$^{+0.03}_{-0.03}$ &13.4$^{+1.6}_{-1.6}$ &268.5$^{+8.6}_{-7.3}$ &103.0$^{+1.3}_{-1.6}$ &-1.97$^{+0.03}_{-0.02}$ &7.5$^{+1.1}_{-1.1}$ &254.9 &$\frac{240.6}{234}$\\\hline
[1.2, 1.7] &mBB &0.35$^{+0.01}_{-0.01}$ &1.8$^{+0.9}_{-0.5}$ &418.3$^{+6.0}_{-5.6}$ &232.3$^{+2.7}_{-1.6}$ &  &  &222.2 &$\frac{212.7}{236}$\\
& &0.38$^{+0.03}_{-0.02}$ &8.5$^{+2.1}_{-1.4}$ &413.2$^{+6.7}_{-6.2}$ &229.4$^{+3.7}_{-3.1}$ &-1.95$^{+0.05}_{-0.03}$ &6.0$^{+1.0}_{-1.1}$ &228.9 &$\frac{214.6}{234}$\\\hline
[1.7, 2.2] & &0.53$^{+0.01}_{-0.01}$ &1.6$^{+0.6}_{-0.4}$ &462.9$^{+5.5}_{-4.7}$ &382.0$^{+4.3}_{-2.5}$ &  &  &333.6 &$\frac{324.1}{236}$\\
&mBB+PL &0.64$^{+0.03}_{-0.03}$ &9.0$^{+4.1}_{-3.2}$ &445.3$^{+6.7}_{-20.0}$ &374.7$^{+3.6}_{-3.6}$ &-1.95$^{+0.03}_{-0.05}$ &8.4$^{+2.0}_{-2.5}$ &302.1 &$\frac{287.8}{234}$\\\hline
[2.2, 4.6]& &0.18$^{+0.02}_{-0.02}$ &10.0$^{+7.3}_{-0.7}$ &439.1$^{+127.0}_{-15.7}$ &270.3$^{+115.9}_{-6.3}$ &  &  &269.1 &$\frac{259.7}{220}$\\
&mBB+PL &0.21$^{+0.03}_{-0.03}$ &11.0$^{+2.2}_{-1.8}$ &431.4$^{+20.0}_{-15.0}$ &263.1$^{+7.5}_{-7.5}$ &-1.98$^{+0.20}_{-0.11}$ &6.9$^{+10.0}_{-5.5}$ &268.7 &$\frac{254.6}{218}$\\\hline
[4.6, 7.0]& &0.14$^{+0.03}_{-0.04}$ &10.0$^{+4.8}_{-0.6}$ &492.0$^{+33.0}_{-16.9}$ &394.6$^{+15.6}_{-8.6}$ &  &  &283.5 &$\frac{274.1}{220}$\\
&mBB+PL &0.16$^{+0.10}_{-0.04}$ &12.5$^{+1.5}_{-1.5}$ &493.4$^{+126.0}_{-21.0}$ &385.9$^{+13.6}_{-17.8}$ &-1.98$^{+0.37}_{-0.05}$ &15.8$^{+13.0}_{-10.4}$ &282.9 &$\frac{268.8}{218}$\\\hline
[7.0, 9.3]& &-0.03$^{+0.03}_{-0.19}$ &10.5$^{+3.3}_{-0.6}$ &507.3$^{+110.0}_{-22.1}$ &271.0$^{+73.0}_{-6.1}$ &  &  &311.6 &$\frac{302.2}{220}$\\
&mBB+PL &-0.02$^{+0.03}_{-0.04}$ &11.4$^{+1.9}_{-0.9}$ &506.4$^{+40.0}_{-20.0}$ &265.4$^{+5.9}_{-5.9}$ &-1.98$^{+0.20}_{-0.40}$ &8.0$^{+8.5}_{-7.7}$ &310.6 &$\frac{296.5}{218}$\\\hline
[9.3, 11.7]&&-0.23$^{+0.03}_{-0.21}$ &10.7$^{+1.5}_{-0.5}$ &491.8$^{+110.0}_{-30.3}$ &177.0$^{+60.1}_{-3.1}$ &  &  &270.7 &$\frac{261.3}{220}$\\ 
&mBB+PL &-0.22$^{+0.03}_{-0.04}$ &11.5$^{+4.1}_{-0.8}$ &490.7$^{+80.0}_{-52.1}$ &171.5$^{+11.1}_{-8.4}$ &-1.99$^{+0.20}_{-0.32}$ &9.4$^{+10.0}_{-7.6}$ &267.7 &$\frac{253.6}{218}$\\\hline\hline
[2.2, 11.7] & &0.03$^{+0.01}_{-0.01}$ &10.1$^{+0.4}_{-0.3}$ &504.4$^{+13.6}_{-10.5}$ &288.2$^{+4.2}_{-3.8}$ &  &  &340.0 &$\frac{330.6}{220}$\\
&mBB+PL &0.04$^{+0.02}_{-0.05}$ &10.9$^{+25.1}_{-0.8}$ &502.9$^{+12.0}_{-10.0}$ &281.9$^{+4.1}_{-4.1}$ &-2.03$^{+0.43}_{-0.15}$ &12.8$^{+10.0}_{-10.0}$ &319.3 &$\frac{305.2}{218}$\\\hline\hline
[11.7, 12.2] & &-0.18$^{+0.02}_{-0.02}$ &6.5$^{+0.4}_{-0.4}$ &454.2$^{+12.9}_{-12.1}$ &151.8$^{+2.0}_{-2.0}$ &  &  &298.2 &$\frac{288.6}{236}$\\
 &mBB+PL &-0.19$^{+0.01}_{-0.03}$ &9.8$^{+0.7}_{-0.7}$ &461.5$^{+10.0}_{-10.0}$ &146.2$^{+1.9}_{-1.9}$ &-1.98$^{+0.02}_{-0.02}$ &9.3$^{+1.5}_{-1.7}$ &291.8 &$\frac{277.5}{234}$\\\hline
[12.2, 12.7]& &-0.14$^{+0.02}_{-0.02}$ &6.1$^{+0.4}_{-0.4}$ &486.2$^{+12.1}_{-11.9}$ &160.0$^{+2.0}_{-2.0}$ &  &  &258.3 &$\frac{248.8}{236}$\\
&mBB+PL &-0.15$^{+0.02}_{-0.02}$ &9.7$^{+0.7}_{-0.7}$ &491.5$^{+13.2}_{-12.6}$ &154.6$^{+2.4}_{-2.4}$ &-1.97$^{+0.01}_{-0.01}$ &9.3$^{+1.7}_{-2.0}$ &256.2 &$\frac{241.9}{234}$\\\hline
[12.7, 13.2] & &-0.17$^{+0.02}_{-0.02}$ &6.5$^{+0.3}_{-0.3}$ &409.2$^{+9.9}_{-9.6}$ &178.4$^{+2.0}_{-2.3}$ &  &  &305.4 &$\frac{295.8}{236}$\\
&mBB+PL &-0.20$^{+0.02}_{-0.02}$ &10.2$^{+0.5}_{-0.9}$ &419.8$^{+10.6}_{-10.5}$ &171.7$^{+2.6}_{-2.6}$ &-1.98$^{+0.02}_{-0.01}$ &11.9$^{+1.8}_{-1.7}$ &290.8 &$\frac{276.5}{234}$\\\hline
[13.2, 13.7]&  &-0.23$^{+0.01}_{-0.01}$ &5.7$^{+0.3}_{-0.3}$ &469.9$^{+13.6}_{-12.7}$ &148.9$^{+2.0}_{-1.9}$ &  &  &299.0 &$\frac{289.5}{236}$\\
&mBB+PL &-0.23$^{+0.02}_{-0.01}$ &7.8$^{+0.9}_{-0.9}$ &470.1$^{+14.9}_{-16.2}$ &143.3$^{+2.9}_{-2.7}$ &-1.97$^{+0.02}_{-0.02}$ &8.6$^{+2.1}_{-2.3}$ &298.8 &$\frac{284.6}{234}$\\\hline
[13.7, 14.2]& &-0.30$^{+0.02}_{-0.02}$ &5.8$^{+0.3}_{-0.4}$ &433.1$^{+19.5}_{-13.9}$ &106.3$^{+1.2}_{-2.3}$ &  &  &238.8 &$\frac{229.3}{236}$\\
&mBB+PL&-0.31$^{+0.02}_{-0.02}$ &9.8$^{+0.8}_{-0.8}$ &441.0$^{+16.8}_{-16.0}$ &99.3$^{+1.3}_{-1.3}$ &-1.98$^{+0.04}_{-0.01}$ &12.2$^{+1.5}_{-1.6}$ &215.5 &$\frac{201.2}{234}$\\\hline
[14.2, 14.7] & &-0.34$^{+0.02}_{-0.01}$ &5.5$^{+0.3}_{-0.2}$ &442.3$^{+17.6}_{-16.6}$ &105.4$^{+2.0}_{-2.0}$ &  &  &260.7 &$\frac{251.2}{236}$\\
&mBB+PL&-0.35$^{+0.03}_{-0.02}$ &9.0$^{+0.9}_{-0.7}$ &449.7$^{+23.8}_{-15.8}$ &97.9$^{+2.7}_{-1.4}$ &-1.99$^{+0.02}_{-0.01}$ &13.1$^{+1.7}_{-1.7}$ &240.3 &$\frac{226.0}{234}$\\\hline
[14.7, 15.2] & &-0.36$^{+0.02}_{-0.02}$ &5.6$^{+0.3}_{-0.3}$ &415.5$^{+15.8}_{-15.1}$ &103.5$^{+1.2}_{-1.2}$ &  &  &240.7 &$\frac{231.2}{234}$\\
 &mBB+PL &-0.36$^{+0.02}_{-0.03}$ &7.0$^{+0.8}_{-0.7}$ &417.2$^{+17.1}_{-16.0}$ &99.6$^{+1.3}_{-2.7}$ &-2.03$^{+0.03}_{-0.02}$ &7.0$^{+2.6}_{-2.4}$ &239.4 &$\frac{225.2}{232}$\\\hline
[15.2, 15.7] & &-0.34$^{+0.02}_{-0.02}$ &5.0$^{+0.2}_{-0.5}$ &352.2$^{+14.6}_{-14.6}$ &82.1$^{+1.4}_{-1.4}$ &  &  &271.0 &$\frac{261.5}{234}$\\
&mBB+PL&-0.34$^{+0.02}_{-0.03}$ &7.7$^{+0.4}_{-0.8}$ &353.9$^{+15.1}_{-15.4}$ &76.1$^{+1.9}_{-1.4}$ &-1.98$^{+0.02}_{-0.01}$ &10.4$^{+1.8}_{-2.1}$ &262.0 &$\frac{247.7}{232}$\\\hline
 &BAND   &$\alpha$ &$\beta$ &$E_{\rm peak}$ &  &  &  & & \\\hline
 [-0.01, 0.39]&BAND &-0.77$^{+0.02}_{-0.02}$ &-7.43$^{+2.80}_{-0.47}$ &180.7$^{+4.3}_{-4.4}$& &  &31.3$^{+0.6}_{-0.3}$ &243.30 &$\frac{234.0}{207}$\\\hline
[0.39, 0.7]&BAND &-1.56$^{+0.07}_{-0.11}$ &-6.88$^{+3.66}_{-3.49}$ &98.3$^{+17.0}_{-26.5}$& &  &5.8$^{+0.9}_{-0.6}$ &126.59 &$\frac{117.9}{147}$\\\hline
[0.7, 1.2]&BAND &-0.52$^{+0.03}_{-0.00}$ &-8.98$^{+3.62}_{-1.50}$ &599.9$^{+10.3}_{-2.4}$& &  &109.0$^{+0.10}_{-0.90}$ &248.78 &$\frac{239.3}{236}$\\\hline
[1.2, 1.7]&BAND &-0.49$^{+0.01}_{-0.03}$ &-11.94$^{+2.00}_{-3.58}$ &968.9$^{+8.5}_{-7.2}$& &  &233.2$^{+1.8}_{-2.4}$ &265.97 &$\frac{256.5}{236}$\\\hline
[1.7, 2.2]&BAND &-0.31$^{+0.01}_{-0.03}$ &-9.40$^{+2.90}_{-9.87}$ &1117.4$^{+17.1}_{-10.0}$& &  &382.5$^{+6.2}_{-3.6}$ &417.51 &$\frac{408.0}{236}$\\\hline
[2.2, 4.575] &BAND &-0.61$^{+0.02}_{-0.02}$ &-7.28$^{+3.13}_{-3.00}$ &998.4$^{+41.7}_{-34.0}$ &  & &278.3$^{+5.1}_{-6.6}$ &280.20 &$\frac{270.8}{220}$\\\hline
[4.575, 6.95]&BAND &-0.67$^{+0.01}_{-0.02}$ &-7.36$^{+10.34}_{-3.00}$ &1115.0$^{+26.1}_{-26.1}$ &  & &412.1$^{+4.6}_{-8.2}$ &297.46 &$\frac{288.1}{220}$\\\hline
[6.95, 9.325]&BAND&-0.80$^{+0.01}_{-0.02}$ &-7.21$^{+3.10}_{-3.00}$ &1037.8$^{+44.5}_{-29.7}$ &  & &279.9$^{+9.8}_{-7.5}$ &330.89 &$\frac{321.5}{220}$\\\hline
[9.325, 11.7] &BAND &-0.94$^{+0.02}_{-0.02}$ &-6.70$^{+0.13}_{-0.13}$ &838.3$^{+38.0}_{-44.0}$ &  & &178.92$^{+5.3}_{-10.0}$ &295.42 &$\frac{286.0}{220}$\\\hline
\enddata
\end{deluxetable*}

 \begin{deluxetable*}{lccccccccc}
\centering{Table. \ref{tab:fitresTR}--- Continued. }
\tablehead{
\colhead{Time bins}
& Favoured or Used Model
&\colhead{$\alpha$}
&\colhead{$\beta$}
&\colhead{$E_{\rm peak}$}
&\colhead{}
&\colhead{}
&\colhead{$F_{\rm NT}$}
&\colhead{BIC}
&\colhead{$\frac{\chi^2}{ndof}$}\\
\colhead{(s)}
&\colhead{}
&\colhead{}
&\colhead{}
&\colhead{(keV)}
&\colhead{}
&\colhead{}
&\colhead{(10$^{-6}$ erg cm$^{-2}$ s$^{-1}$)}
&\colhead{}
&\colhead{}}
\startdata
[11.7, 12.2]&BAND &-0.92$^{+0.03}_{-0.01}$ &-8.12$^{+3.82}_{-5.67}$ &872.7$^{+76.9}_{-0.0}$& &  &154.3$^{+2.7}_{-0.0}$ &331.96 &$\frac{322.4}{236}$\\\hline
[12.2, 12.7]&BAND &-0.90$^{+0.05}_{-0.02}$ &-9.61$^{+2.32}_{-7.22}$ &939.0$^{+10.6}_{-7.1}$& &  &159.9$^{+2.0}_{-2.5}$ &278.60 &$\frac{269.1}{236}$\\\hline
[12.7, 13.2]&BAND &-0.90$^{+0.01}_{-0.01}$ &-10.87$^{+1.07}_{-8.34}$ &756.7$^{+6.3}_{-11.6}$& &  &180.7$^{+3.3}_{-4.1}$ &315.01 &$\frac{305.5}{236}$\\\hline
[13.2, 13.7]&BAND &-0.99$^{+0.01}_{-0.01}$ &-9.08$^{+3.72}_{-6.08}$ &835.7$^{+10.0}_{-10.0}$& &  &149.2$^{+1.1}_{-2.3}$ &314.46 &$\frac{304.9}{236}$\\\hline
[13.7, 14.2]&BAND &-1.05$^{+0.01}_{-0.01}$ &-10.71$^{+1.23}_{-8.51}$ &742.5$^{+5.2}_{-7.9}$& &  &107.8$^{+1.4}_{-1.4}$ &291.45 &$\frac{281.9}{236}$\\\hline
[14.2, 14.7]&BAND &-1.09$^{+0.01}_{-0.02}$ &-8.61$^{+3.14}_{-6.39}$ &722.0$^{+9.0}_{-5.0}$& &  &106.9$^{+1.2}_{-1.2}$ &254.52 &$\frac{245.0}{236}$\\\hline
[14.7, 15.2]&BAND &-1.08$^{+0.01}_{-0.03}$ &-8.74$^{+3.20}_{-2.46}$ &657.0$^{+12.5}_{-10.4}$& &  &104.2$^{+1.2}_{-1.2}$ &284.62 &$\frac{275.1}{236}$\\\hline
[15.2, 15.7]&BAND &-1.09$^{+0.01}_{-0.02}$ &-8.55$^{+9.56}_{-0.00}$ &586.1$^{+17.0}_{-17.0}$& &  &84.1$^{+1.4}_{-0.7}$ &284.59 &$\frac{275.1}{236}$\\\hline
[15.7, 16.2]&BAND &-1.12$^{+0.01}_{-0.02}$ &-8.61$^{+2.00}_{-2.71}$ &793.3$^{+28.9}_{-19.3}$& &  &89.4$^{+1.8}_{-1.2}$ &281.45 &$\frac{271.9}{236}$\\\hline
[16.2, 16.7]&BAND &-1.07$^{+0.02}_{-0.03}$ &-8.40$^{+2.73}_{-2.98}$ &685.5$^{+20.0}_{-93.3}$& &  &81.3$^{+1.5}_{-1.0}$ &267.13 &$\frac{257.6}{236}$\\\hline
[16.7, 17.2]&BAND &-1.17$^{+0.01}_{-0.02}$ &-8.44$^{+2.37}_{-7.47}$ &481.6$^{+20.0}_{-20.0}$& &  &49.7$^{+0.6}_{-0.6}$ &266.30 &$\frac{256.8}{236}$\\\hline
[17.2, 17.7]&BAND &-1.16$^{+0.06}_{-0.01}$ &-8.27$^{+3.86}_{-2.00}$ &511.5$^{+22.1}_{-14.7}$& &  &48.8$^{+1.2}_{-0.6}$ &199.33 &$\frac{189.8}{236}$\\\hline
[17.7, 18.2]&BAND &-1.29$^{+0.05}_{-0.03}$ &-7.58$^{+0.12}_{-2.10}$ &197.6$^{+28.3}_{-11.6}$& &  &13.1$^{+0.3}_{-0.5}$ &243.51 &$\frac{234.0}{236}$\\\hline
[18.2, 18.7]&BAND &-1.55$^{+0.06}_{-0.03}$ &-7.06$^{+3.06}_{-3.00}$ &219.6$^{+32.4}_{-39.3}$& &  &10.9$^{+0.6}_{-0.8}$ &184.64 &$\frac{175.1}{236}$\\\hline
[18.7, 19.2]&BAND &-1.30$^{+0.03}_{-0.03}$ &-7.75$^{+3.05}_{-2.90}$ &380.0$^{+20.0}_{-20.0}$& &  &30.6$^{+0.6}_{-0.6}$ &204.36 &$\frac{194.8}{236}$\\\hline
[19.2, 19.7]&BAND &-1.23$^{+0.02}_{-0.02}$ &-8.40$^{+2.63}_{-2.00}$ &504.9$^{+20.0}_{-26.1}$& &  &56.8$^{+0.9}_{-0.9}$ &264.82 &$\frac{255.3}{236}$\\\hline
[19.7, 20.2]&BAND &-1.22$^{+0.01}_{-0.01}$ &-8.70$^{+0.87}_{-8.67}$ &715.6$^{+12.5}_{-10.4}$& &  &78.7$^{+1.7}_{-1.1}$ &308.10 &$\frac{298.6}{236}$\\\hline
[20.2, 20.7]&BAND &-1.16$^{+0.03}_{-0.01}$ &-8.48$^{+9.84}_{-3.20}$ &730.0$^{+17.2}_{-9.1}$& &  &70.3$^{+1.4}_{-1.4}$ &292.01 &$\frac{282.5}{236}$\\\hline
[20.7, 21.2]&BAND &-1.12$^{+0.02}_{-0.00}$ &-8.38$^{+9.56}_{-0.25}$ &526.2$^{+25.1}_{-18.8}$& &  &54.6$^{+1.1}_{-1.1}$ &280.33 &$\frac{270.8}{236}$\\\hline
[21.2, 21.7]&BAND &-1.12$^{+0.03}_{-0.00}$ &-8.31$^{+5.21}_{-4.59}$ &593.3$^{+20.0}_{-20.0}$& &  &71.8$^{+1.1}_{-1.1}$ &267.21 &$\frac{257.7}{236}$\\\hline
[21.7, 22.2]&BAND &-1.19$^{+0.02}_{-0.02}$ &-8.51$^{+2.36}_{-7.45}$ &561.1$^{+24.0}_{-24.0}$& &  &51.5$^{+0.9}_{-0.9}$ &239.74 &$\frac{230.2}{236}$\\\hline
[22.2, 22.7]&BAND &-1.21$^{+0.00}_{-0.03}$ &-8.44$^{+3.12}_{-6.50}$ &545.2$^{+20.0}_{-20.0}$& &  &45.6$^{+0.7}_{-0.7}$ &240.01 &$\frac{230.5}{236}$\\\hline
[22.7, 23.2]&BAND &-1.15$^{+0.00}_{-0.03}$ &-7.95$^{+2.86}_{-2.80}$ &405.4$^{+20.0}_{-20.0}$& &  &41.1$^{+0.7}_{-0.7}$ &221.29 &$\frac{211.8}{236}$\\\hline
[23.2, 23.7]&BAND &-1.18$^{+0.00}_{-0.03}$ &-7.73$^{+3.87}_{-3.03}$ &292.0$^{+10.0}_{-10.0}$& &  &28.9$^{+0.7}_{-0.7}$ &226.04 &$\frac{216.5}{236}$\\\hline
[23.7, 24.2]&BAND &-1.26$^{+0.02}_{-0.02}$ &-8.09$^{+9.75}_{-0.00}$ &406.7$^{+16.0}_{-21.4}$& &  &40.2$^{+0.9}_{-0.9}$ &270.98 &$\frac{261.5}{236}$\\\hline
[24.2, 24.7]&BAND &-1.29$^{+0.01}_{-0.01}$ &-8.31$^{+2.72}_{-3.93}$ &505.9$^{+23.4}_{-31.2}$& &  &44.5$^{+0.5}_{-1.1}$ &256.69 &$\frac{247.2}{236}$\\\hline
[24.7, 25.2]&BAND &-1.25$^{+0.02}_{-0.01}$ &-8.00$^{+2.00}_{-2.00}$ &526.2$^{+21.2}_{-28.2}$& &  &44.5$^{+1.2}_{-1.0}$ &230.45 &$\frac{220.9}{236}$\\\hline
[25.2, 25.7]&BAND &-1.31$^{+0.03}_{-0.02}$ &-8.25$^{+3.36}_{-6.46}$ &340.8$^{+16.0}_{-16.0}$& &  &41.8$^{+0.7}_{-0.7}$ &266.04 &$\frac{256.5}{236}$\\\hline
[25.7, 26.2]&BAND &-1.37$^{+0.02}_{-0.03}$ &-7.97$^{+9.83}_{-0.00}$ &271.0$^{+12.3}_{-9.3}$& &  &31.9$^{+0.7}_{-0.4}$ &228.73 &$\frac{219.2}{236}$\\\hline
[26.2, 26.7]&BAND &-1.48$^{+0.02}_{-0.02}$ &-7.20$^{+4.24}_{-2.10}$ &192.4$^{+15.1}_{-10.8}$& &  &24.1$^{+0.5}_{-0.5}$ &195.13 &$\frac{185.6}{236}$\\\hline
[26.7, 27.2]&BAND &-1.61$^{+0.03}_{-0.03}$ &-7.11$^{+3.20}_{-3.30}$ &240.7$^{+66.7}_{-32.7}$& &  &25.4$^{+1.0}_{-1.0}$ &259.15 &$\frac{249.6}{236}$\\\hline
[27.2, 27.7]&BAND &-1.54$^{+0.03}_{-0.01}$ &-7.69$^{+2.10}_{-3.60}$ &404.0$^{+8.1}_{-13.2}$& &  &26.4$^{+0.9}_{-0.6}$ &240.60 &$\frac{231.1}{236}$\\\hline
[27.7, 28.2]&BAND &-1.44$^{+0.01}_{-0.03}$ &-8.28$^{+0.37}_{-9.50}$ &453.9$^{+10.6}_{-4.1}$& &  &37.3$^{+0.6}_{-1.2}$ &203.75 &$\frac{194.2}{236}$\\\hline
[28.2, 28.7]&BAND &-1.41$^{+0.03}_{-0.01}$ &-7.76$^{+1.20}_{-2.08}$ &528.9$^{+21.1}_{-10.4}$& &  &40.7$^{+0.9}_{-0.9}$ &233.97 &$\frac{224.4}{236}$\\\hline
[28.7, 29.2]&BAND &-1.44$^{+0.00}_{-0.02}$ &-7.38$^{+0.80}_{-3.20}$ &516.8$^{+45.7}_{-45.7}$& &  &24.2$^{+1.1}_{-0.7}$ &232.61 &$\frac{223.1}{236}$\\\hline
[29.2, 29.7]&BAND &-1.45$^{+0.02}_{-0.02}$ &-7.71$^{+0.70}_{-1.20}$ &394.2$^{+0.0}_{-58.4}$& &  &19.0$^{+0.6}_{-0.8}$ &215.80 &$\frac{206.3}{236}$\\\hline
[29.7, 30.2]&BAND &-1.43$^{+0.01}_{-0.03}$ &-8.26$^{+9.87}_{-0.00}$ &510.1$^{+43.1}_{-43.1}$& &  &24.9$^{+1.0}_{-0.8}$ &220.93 &$\frac{211.4}{236}$\\\hline
[30.2, 30.7]&BAND &-1.33$^{+0.03}_{-0.03}$ &-7.16$^{+2.10}_{-1.20}$ &266.1$^{+6.0}_{-2.0}$& &  &13.2$^{+0.4}_{-0.6}$ &222.34 &$\frac{212.8}{236}$\\\hline
[30.7, 31.2]&BAND &-1.38$^{+0.03}_{-0.03}$ &-7.22$^{+3.00}_{-2.00}$ &211.0$^{+7.2}_{-3.2}$& &  &11.6$^{+0.4}_{-0.4}$ &196.71 &$\frac{187.2}{236}$\\\hline
[31.2, 31.7]&BAND &-1.40$^{+0.06}_{-0.06}$ &-7.18$^{+3.00}_{-3.12}$ &149.2$^{+12.0}_{-19.7}$& &  &9.4$^{+0.4}_{-0.4}$ &177.98 &$\frac{168.5}{236}$\\\hline
[31.7, 35.7]&BAND &-1.55$^{+0.02}_{-0.02}$ &-7.98$^{+5.42}_{-1.69}$ &254.5$^{+9.0}_{-9.0}$& &  &14.9$^{+0.2}_{-0.2}$ &300.80 &$\frac{291.3}{236}$\\\hline
[35.7, 39.7]&BAND &-1.51$^{+0.02}_{-0.00}$ &-7.89$^{+9.76}_{-0.00}$ &192.8$^{+8.2}_{-10.1}$& &  &10.4$^{+0.1}_{-0.1}$ &255.69 &$\frac{246.2}{236}$\\\hline
[39.7, 43.7]&BAND &-1.76$^{+0.02}_{-0.02}$ &-7.10$^{+3.00}_{-3.40}$ &125.3$^{+10.0}_{-10.0}$& &  &7.5$^{+1.1}_{-1.0}$ &185.46 &$\frac{175.9}{236}$\\\hline
[43.7, 47.7]&BAND &-1.67$^{+0.02}_{-0.02}$ &-7.49$^{+3.00}_{-3.00}$ &162.8$^{+9.1}_{-13.8}$& &  &6.7$^{+0.1}_{-0.5}$ &176.36 &$\frac{166.8}{236}$\\\hline
[47.7, 51.7]&BAND &-1.54$^{+0.02}_{-0.02}$ &-7.19$^{+2.40}_{-1.22}$ &142.7$^{+11.1}_{-4.9}$& &  &4.8$^{+0.0}_{-0.2}$ &186.30 &$\frac{176.8}{236}$\\\hline
[51.7, 61.7]&BAND &-1.67$^{+0.03}_{-0.03}$ &-7.64$^{+3.99}_{-5.74}$ &88.1$^{+12.1}_{-6.5}$& &  &3.5$^{+0.2}_{-0.2}$ &163.82 &$\frac{154.3}{236}$\\\hline
[61.7, 71.7]&BAND &-1.70$^{+0.03}_{-0.05}$ &-7.22$^{+3.00}_{-0.12}$ &98.3$^{+8.0}_{-7.0}$& &  &2.4$^{+0.0}_{-0.0}$ &168.22 &$\frac{158.7}{236}$\\\hline
[71.7, 81.7]&BAND &-1.77$^{+0.20}_{-0.10}$ &-6.92$^{+3.00}_{-0.12}$ &43.2$^{+10.0}_{-12.0}$& &  &1.3$^{+0.5}_{-0.2}$ &143.93 &$\frac{134.4}{236}$\\\hline
[81.7, 91.7]&BAND &-1.79$^{+0.19}_{-0.10}$ &-7.22$^{+5.49}_{-2.70}$ &46.5$^{+10.0}_{-10.0}$& &  &1.3$^{+0.1}_{-0.4}$ &157.30 &$\frac{147.8}{236}$\\\hline
\enddata
\end{deluxetable*}

\clearpage
\bibliography{GRB230307A}{}
\bibliographystyle{aasjournal}


\end{document}